\documentclass[12pt]{article}
\usepackage{graphicx}
\usepackage{amsmath}
\usepackage{vector}
\usepackage{pdfpages}
\usepackage{graphics}
\usepackage{xypic}
\usepackage[cm]{fullpage}
\usepackage{bbm}
\usepackage{graphicx}
\usepackage{graphics}
\usepackage{fancybox}
\usepackage{amsmath,float,latexsym,psfrag,epsf,epsfig,amssymb,
chicago,rotating}
\usepackage{color}
\usepackage{url}
\usepackage{fullpage}
\usepackage[ruled,linesnumbered]{algorithm2e}
\usepackage{caption}
\usepackage{subcaption}

\newcommand{\K}{{\mathbf{K}}}

\newcommand{\z}{{\mathbf{z}}}

\newcommand{\Y}{{\mathbf{Y}}}

\newcommand{\Z}{{\mathbf{Z}}}

\newcommand{\I}{{\mathbf{I}}}

\newcommand{\bftheta}{\hbox{\boldmath$\theta$}}

\newcommand{\bftau}{\hbox{\boldmath$\tau$}}
\newcommand{\bfnu}{\hbox{\boldmath$\nu$}}
\newcommand{\bfsigma}{\hbox{\boldmath$\sigma$}}
\newcommand{\bflambda}{\hbox{\boldmath$\lambda$}}

\newcommand{\bfmu}{\hbox{\boldmath$\mu$}}

\begin{document}

% \maketitle
%  \vspace{-.75in}
% \begin{center}
% \date 
% \end{center}

\begin{center}

{\Large \bfseries Bayesian model selection for the latent position cluster model for Social Networks}
\vspace{5 mm}

{\large N. FRIEL$^{1,2}$, C. RYAN$^{1}$ and J. WYSE$^{1,2}$}\\ %\footnote{\texttt{nial.friel@ucd.ie}}} \\
{\textit{$^1$School of Mathematical Sciences and $^2$Insight research centre, \\ University College Dublin, Ireland.}}
% $^3$School of Computer Science and Statistics, Trinity College Dublin, Ireland.}}
\vspace{5 mm}

\today 

\vspace{5mm}

\end{center}

\begin{center}
\begin{abstract}

The latent position cluster model is a popular model for the statistical analysis of network data. This approach assumes that 
there is an underlying latent space in which the actors follow a finite mixture distribution. Moreover, actors which are close in 
this latent space tend to be tied by an edge. This is an appealing approach since it allows the model to cluster actors which 
consequently provides the practitioner with useful qualitative information. However, exploring the uncertainty in the number 
of underlying latent components in the mixture distribution is a very complex task. The current state-of-the-art is to use an
approximate form 
of BIC for this purpose, where an approximation of the log-likelihood is used instead of the true log-likelihood which is unavailable. 
The main contribution of this paper is to show that through the use of conjugate prior distributions it is possible to analytically 
integrate out almost all of the model parameters, leaving a posterior distribution which depends on the allocation vector of the mixture model. A
consequence of this is that it is possible to carry out posterior inference over the number of components in the latent mixture
distribution without using trans-dimensional MCMC algorithms such as reversible jump MCMC. Moreover, our algorithm allows for more reasonable 
computation times for larger networks than the standard methods using the \texttt{latentnet} package \cite{Kriv:Hand07,Kriv:Hand13}.

\end{abstract}

\end{center}

\section{Introduction}

A social network consists of nodes or actors in a graph, for example, individuals or organizations, 
connected by one or more specific types of interdependency, such as, friendship, business relationships or trade between 
countries. 
% The resulting networks are typically complex and provide a challenge for statisticians to model and provide
% useful inference. %Maybe improve the last sentence... 
The analysis of network data has a rich interdisciplinary history finding application in a wide range of areas including 
sociology \cite{wasserman:galaskiewicz1994}, 
physics \shortcite{adamicetal01}, 
biology \cite{michailidis12}, 
computer science \shortcite{faloutsosetal1999}
and many more. The aims of network analysis are both descriptive and inferential.
For example, one might be interested in examining global structure within a network or in analysing network attributes such 
as the degree distribution as well as the local structure such as the identification of influential or highly connected actors 
in the network. Inferential goals include hypothesis testing, model comparison and making predictions, for example, how far
will a virus spread through a network.

There have been many statistical models proposed for the analysis of network data, the most popular of which include the 
exponential random graph model see Wasserman and Pattison \citeyear{wasserman:pattison1996} and Robins \textit{et al} \citeyear{robinsetal07b}
 and the stochastic block model of Nowicki and Snijders \citeyear{nowicki:snijders01} and it's variants. 
For a recent perspective on the statistical analysis of network data, see Kolaczyk \citeyear{kolaczyk09}.
An alternative and popular approach to modelling network data is the latent space approach \shortcite{hoff:raft:hand02}. 

Here each actor is embedded in a latent `social space' in which actors that are close in the latent space are more likely to be tied by an edge. 
Latent space models  \shortcite{handcocketal07} naturally accommodate many sociological features such as homophily, reciprocity 
and transitivity.
The recent development of  Handcock \textit{et al} \citeyear{handcocketal07} extends the latent space model of Hoff \textit{et al} \citeyear{hoff:raft:hand02}
 to cluster actors directly, where the positions of actors are assumed to be distributed according
to a finite mixture. 
The latent position cluster model provides a useful interpretation of the network since the underlying latent model 
provides an automatic means of clustering actors while also providing the uncertainty around the probability of actor membership
to each cluster. The R package \texttt{latentnet} \cite{Kriv:Hand07,Kriv:Hand13}, which is part of the \texttt{statnet} 
suite of packages, can be used to fit the latent position cluster model.

Despite its popularity, a major difficulty with the latent position cluster model is inferring the number of components in the 
latent mixture distribution. The approach advocated by Handcock \textit{et al} \citeyear{handcocketal07} is to assess this uncertainty by estimating the Bayesian
information criterion (BIC) for each possible model. However, it turns out that it is computationally prohibitive to calculate the maximum log likelihood
used in BIC and a tractable approximation is to condition on the minimum Kullback-Leibler  estimate of the actors latent positions \shortcite{shortreedetal06},
rather than integrating over the posterior distribution of the actors positions, thereby accounting for the uncertainty in
these latent positions. Note that a variational Bayes approximation has been proposed by Salter-Townshend and Murphy 
\citeyear{salter:murphy12} but it too uses the same strategy as \shortcite{handcocketal07} to infer the number of components.
One of the primary contributions of this article is to resolve this issue. To this end we use conjugate prior distributions which
allow almost all latent mixture parameters to be integrated out. This results in a collapsed posterior distribution which depends on the
vector of allocations of actors to components. The important consequence of this is that 
% WHAT ABOUT EMPTY GROUPS! the allocation vector encodes the number of components of the mixture distribution, but crucially, 
the number of components can be inferred without the use of 
trans-dimensional MCMC techniques such as reversible jump Markov chain Monte Carlo \cite{richardson:green97}. This approach is similar 
to that presented in Nobile and Fearnside \citeyear{nob:fearn07} and Wyse and Friel \citeyear{wyse:friel12} for the collapsed finite mixture model and
latent block models, respectively. The second important 
contribution of this paper is that our approach is computationally fast and can be applied to larger networks than is feasible using \texttt{latentnet}. 
The software which
accompanies this paper can be used to implement all the examples presented herein.

The paper begins in Section \ref{lpcm} by describing the latent position cluster model and the current approach to inferring the number of clusters.
Section \ref{collapsing} introduces the collapsed form of the model. 
Cross-model inference is described for the collapsed latent position cluster model in Section \ref{algs}. 
Section \ref{res} applies and compares the methodology to current methods for some known social network data. 
Some discussions follow in Section \ref{discussion}.
 
\section {The Latent Position Cluster Model for Social Networks}\label{lpcm}

Network data may be represented by an $n \times n$ adjacency matrix 
% $\Y=\{y_{ij}:i,\ldots,n; j=1,\ldots,n\}$
$\Y=\{y_{ij}\}_{i,j=1}^n$  
of binary relations $y_{ij}$ between actors $i$ and $j$, 
indicating presence or absence of a tie between $i$ and $j$. Ties can be directed where $y_{ij}$ does not necessarily equal to $y_{ji}$ or undirected 
where $y_{ij} = y_{ji}$ for all $i\neq j$. %depending on the reciprocal nature of the relationship between actors. 
Self-ties are typically not allowed thus diagonal entries of this matrix take the value $0$. It is also possible to consider
networks with integer or weighted values representing the strength of relationship between the two connected actors. 

The Latent Position Cluster Model was introduced by Handcock \textit{et al} \citeyear{handcocketal07} where actor $i$ is assumed to have an unobserved random position, $\z_i$, 
in a $d$-dimensional Euclidean latent social space. The choice of $d=2$ aids visualization but the latent space could be of any dimension with the possibility to infer
$d$ as a parameter in the model.
The probability of a link between two actors is assumed independent of all other links in the network, given the latent locations
%  $\Z=\{\z_{i}:i,\ldots,n\}$ 
 $\Z=\{\z_{i}\}_{i=1}^n$ 
of the actors, resulting in the likelihood
\begin{equation}
 L(\Y|\Z,\beta) = \prod_{i\neq j} \pi(y_{ij}|\z_i,\z_j,\beta).
\label{eqn:likelihood}
\end{equation}
A logistic regression model is employed, where the probability of a tie between actors $i$ and $j$ 
depends on the Euclidean distance between $\z_i$ and $\z_j$ in the latent social space, $||\z_i - \z_j||$,   
\begin{eqnarray}\label{lpcmeq}
\log{\left \{ \frac{\pi(y_{ij}=1|\z_i,\z_j,\beta)}{\pi(y_{ij}=0|\z_i,\z_j,\beta)} \right \}}=\beta-||\z_i - \z_j||,&& \forall  i,j \in\{1,\dots,n\}, i\neq j, 
\end{eqnarray}
where $\beta$ is an intercept parameter. 
It is assumed that the latent locations $\Z$ are drawn from a finite mixture of $G$ Multivariate Normal components which models the clustering of actors.
The mixture model is
 \begin{equation*}
   \pi(\Z|\bfmu, \bfsigma^2,\bflambda,G)= \prod_{i=1}^n \left( \sum_{g=1}^{G}  \lambda_g  f(\z_i; \bfmu_{g}, \sigma_{g}^2 \I_d)\right),
 \end{equation*}
where $f(\z_i; \bfmu_{g}, \sigma_{g}^2\I_d)$ is the density function of a Multivariate Normal distribution with cluster means $\bfmu=(\bfmu_1,\dots,\bfmu_G)$ and
cluster covariance matrices $\sigma_g^2 \I_d$ for $g=1,\ldots,G$, where $\I_d$ is the $d$-dimensional identity matrix. 
The mixing weights are $\bflambda =(\lambda_1,\dots,\lambda_G)$, where $\lambda_g$ is the probability of actor $i$ belonging to cluster 
$g \in (1,\ldots,G)$ and $\sum_{g} \lambda_g =1$.
% AND/OR??
% The position of actor $i$ is drawn from a multivariate normal distribution; 
% \begin{equation*}
%  z_i \sim \sum_{g=1}^G \lambda_g f(\bfmu_g,\sigma_g^2 \I_d),
% \end{equation*}
% \begin{equation*}
%  \pi(z_i|\bfmu,\sigma^2,\lambda_g,G) = \sum_{g=1}^G \lambda_g f(z_i|\bfmu_g,\sigma_g^2 \I_d).
% \end{equation*}
As usual in mixture modelling, a latent allocation vector $\K=(k_1,\ldots,k_n)$ is introduced where $k_i\in \{1,\ldots,G\}$ for 
all $i=1,\ldots,n$. If actor $i$ belongs to cluster $g$ then $k_i=g$. This provides a tractable augmented expression for the mixture model,
 \begin{equation*}
   \pi(\Z,\K|\bfmu, \bfsigma^2,\bflambda,G)= \prod_{i=1}^n  \prod_{g=1}^{G} \left( \lambda_g  f(\z_i; \bfmu_{g}, \sigma_{g}^2 \I_d)\right)^{\mathbbm{1}(k_i=g)},
 \end{equation*}
where the indicator function ${\mathbbm{1}(k_i=g)}$ is $1$ if $k_i=g$ or $0$ otherwise.
This joint density of $\Z$ and $\K$ can be factorised as 
\begin{equation}
   \pi(\Z|\K, \bfmu, \bfsigma^2,G)\pi(\K|\lambda,G) = \prod_{i=1}^n  \prod_{g=1}^{G} f(\z_i; \bfmu_{g}, \sigma_{g}^2 \I_d)^{\mathbbm{1}(k_i=g)}\prod_{g=1}^G\lambda_g^{(\sum_{i=1}^n \mathbbm{1}(k_i=g))}.
\end{equation}
% where,
% \begin{equation}
%    \pi(\K|\lambda,G)=\prod_g^G\lambda_g^{(\sum_{i=1}^n \mathbbm{1}(k_i=g))},
% \end{equation}
% and where,
% \begin{equation}
%    \pi(\Z|\bfmu, \bfsigma^2,G)\pi(\K|\lambda,G)= \prod_{i=1}^n  \prod_{g=1}^{G} f(\z_i; \bfmu_{g}, \sigma_{g}^2 \I_d)^{\mathbbm{1}(k_i=g)}.
% \end{equation}
The posterior distribution can be factorised as 
\begin{displaymath}
 \pi(\Z, {\bfmu}, \bfsigma^2, \bflambda, \beta, \K, G|\Y) \propto L(\Y|\Z,\beta) \pi(\Z|\bfmu, \bfsigma^2,\K,G)\pi(\K | \bflambda,G)\pi(\bflambda|G)
\pi(\bfmu|\bfsigma^2,G)\pi(\bfsigma^2|G) \pi(\beta) \pi(G).
\end{displaymath}
% The likelihood of a dyad $y_{ij}$ depends on how far apart actors $i$ and $j$ are in the latent social space. The likelihood is defined as:
% \begin{displaymath}
%   L(\Y|\Z,\K,\beta)=\displaystyle\prod_{i=1}^n{}\displaystyle\prod_{j \neq i} \pi(y_{ij}|\z_i,\z_j, \beta)
%  = \displaystyle\prod_{i=1}^n{} \displaystyle\prod_{j \neq i} \frac{\exp\left\{y_{ij}\left(\beta-||\z_i - \z_j||\right)\right\}}{1+\exp\left\{\beta-||\z_i - \z_j||\right\}}.
% \end{displaymath}
% This equation
The likelihood as defined by equations (\ref{eqn:likelihood}) and (\ref{lpcmeq}) involves the product of all possible  $n \times (n-1)$ pairs of actors.
% A likelihood approximation to avoid this computational bottleneck using a case-control sampling approach from  has been developed
% by 
Raftery \textit{et al} \citeyear{rafteryetal2012} adopt the epidemiological approach of case-control sampling \cite{breslow1996} to approximate the likelihood thus 
reducing computation from $O(n^2)$ to $O(n)$.
% Case-control studies in epidemiology compare a group having a disease (cases) to a control group. 
% Since cases are rare, available cases are collected and corresponding controls are sampled from the disease-free population.
% In a network setting, the analogy is that a case is an interaction $y_{ij}=1$ and a control is a non tie $y_{ij}=0$. A simple random sample of non-ties can be chosen
% to match the cases or sampling can be stratified by `closeness' of edges measured by  shortest path length. The shortest path between actors $i$ and $j$
%  is a path with the minimal number of vertices. 
Our analysis does not approximate the likelihood. Instead computation time is reduced
by integrating out the  clustering parameters, $\bftheta=(\bfmu,\bfsigma^2,\bflambda)$.

Prior distributions on the model parameters $\beta$, $\bfmu, \bftau=\frac{1}{\bfsigma^2}$ and $\bflambda$ are 
\begin{eqnarray*} 
 \beta \sim \mathrm{Normal}(\xi, \psi),  	&& \bfmu_g|\tau_g \sim \mathrm{MVN}_d\left(0,\frac{\omega^2}{\tau_g} \I_d\right),\\
 \tau_g=1/\sigma_g^2  \sim \mathrm{Gamma}\left(\frac{\alpha}{2},\frac{\delta}{2}\right), && \bflambda \sim \mathrm{Dirichlet}\left(\bfnu\right),  \\	
\end{eqnarray*}
 where the prior hyper-parameters $\phi=\left(\xi, \psi,\alpha,\delta,\bfnu,\omega^2\right) $ are user specified.
The priors model dependency on the latent positions and their hyperparameter values require careful choice. See a discussion of this point in the rejoinder to the article
\shortcite{handcocketal07}. 
% The choice of priors here is not truly Bayesian in the sense that instead of using expert guided priors, conjugate priors are used to aid mathematics. The advantage of this is this fast and 
% principled approach to model choice. However one must be aware that the choice of priors can have a serious impact on the final results of our analysis.
% $\bfmu_g \sim \mathrm{MVN}_d\left(0,\omega^2 \I_d\right)$.
Following Handcock \textit{et al} \citeyear{handcocketal07}, $\xi=0$ and $\psi=2$ allowing a wide range of values for $\beta$. 
The prior hyper-parameters for the mixing weights $\bflambda$ are fixed as in Handcock \textit{et al} \citeyear{handcocketal07}, where $\nu_g=3$ for $g = 1, \ldots, G$ to put
low probability on small group sizes. 
% Richardson \& Green \cite{richardson:green97} use a value of $\nu=1$. We have experimented with both in sensitivity analysis of hyper parameters.
The prior on the number of clusters is  $\mathrm{Poisson}(1)$ distributed  following Nobile and Fearnside \citeyear{nob:fearn07} which penalizes the addition of 
empty groups. Handcock \textit{et al} \citeyear{handcocketal07} and Richardson \& Green \citeyear{richardson:green97} instead employ a Uniform prior distribution between $1$ and a pre-specified integer $G$.
However, in a technical report, Nobile \citeyear{nobile07} argues that there is a significant effect on the posterior distribution of $G$ from models with empty components. 
The use of a Poisson prior reduces this effect, and has been used by other authors, including Phillips and Smith \citeyear{phil:smith96} and Stephens \citeyear{stephens00a}.
% LEAVE UNLESS ASKED BY REVIEWER 
The conditioning of $\bfmu_g$ on $\tau_g$ is a fully conjugate prior and is commonly employed in the literature
\cite{nob:fearn07,dell:pap06}.
We use this to allow both  parameters to be integrated out of the model
 as in Nobile and Fearnside \citeyear{nob:fearn07}. This differs from the original specification of the model 
by Handcock \textit{et al} \citeyear{handcocketal07},
where the prior distribution for the cluster means is 
\begin{displaymath}
 \bfmu_g \sim \mathrm{MVN}_d\left(\bf{0},\omega^2 \I_d\right),
\end{displaymath}
which does not depend on the cluster variances.
\subsection{Choosing the number of clusters for the latent position cluster model}\label{bicsec}
The model evidence (sometimes called the marginal or integrated likelihood) plays a central role in the Bayesian approach to model choice. 
The model evidence for the latent position cluster model with clustering parameters denoted $\bftheta=(\bfmu,\bfsigma^2,\bflambda)$ and all terms
conditional on $G$ is given by
\begin{eqnarray}
 \pi(\Y|G)  &=& \int_{\Z} \int_{\bftheta} \int_{\beta} \pi(\Y,\Z|\beta,\bftheta)\pi(\beta)\pi(\bftheta)d\beta d\bftheta d\Z \nonumber\\
&=&\int_{\Z} \int_{\bftheta} \int_{\beta}  L(\Y|\Z,\beta) \pi(\Z|\bftheta)\pi(\beta)\pi(\bftheta)d\beta d\bftheta d\Z. \label{eqn:evidence}
\end{eqnarray}
This quantity represents the probability of the observed data given a latent mixture model with $G$ components. Here the variable $G$ can be interpreted as a model
index.
%  The Bayes factor is the ratio of evidence terms and can be used to compare competing models $G_1$ and $G_2$
% \begin{displaymath}
% BF_{G_1,G_2}=\frac{\pi(\Y|G_1)}{\pi(\Y|G_2)}.
% \end{displaymath}

Using the model evidence, Bayes theorem can be used to evaluate the posterior probability, $\pi(G|\Y) \propto \pi(\Y|G)\pi(G)$. 
See Friel and Wyse \citeyear{friel:wyse12} for a recent review of model evidence estimation. 
However, integration across all possible values of $\Z$ in equation (\ref{eqn:evidence}) is intractable, due to the dimensionality of $\Z$. 
A pragmatic approach taken by Handcock \textit{et al} \citeyear{handcocketal07} 
is to condition on a fixed estimate of latent actor locations $\hat\Z$. 
These are estimated using minimum Kullback-Leibler position estimation \shortcite{shortreedetal06}. Since the logistic regression model is a
function of distances between actors rather than the actual latent positions, the estimate $\hat \Z$ is found by minimizing the Kullback-Leibler divergence 
between the true unknown model distances and the MCMC sample position based distances. 
See Appendix A of Handcock \textit{et al} \citeyear{handcocketal07} for further details. % is minimized.
The model evidence is approximated as
\begin{eqnarray}
 \pi(\Y|G)  \approx \pi(\Y,\hat{\Z}|G) &=& \int_{\bftheta} \int_{\beta} \pi(\Y|\hat{\Z},\beta)\pi(\hat{\Z}|\bftheta) \pi(\beta) \pi(\bftheta)d\beta d\bftheta\nonumber\\
&=&  \int_{\beta} \pi(\Y|\hat{\Z},\beta) \pi(\beta)d\beta\int_{\bftheta} \pi(\hat{\Z}|\bftheta) \pi(\bftheta) d\bftheta.\label{eqn:evidenceproduct}
\end{eqnarray}
The `best' $G$ component model corresponds to the largest value of $\pi(\Y,\hat{\Z}|G)$. 
This approach does not take the uncertainty of $\Z$ into account, moreover it is unclear how this approximation impacts upon
the assessment of the number of latent mixture components.
The BIC approximation to the model evidence  \cite{schwarz1978} is employed by Handcock \textit{at al} \citeyear{handcocketal07}  to approximate the integrals in equation (\ref{eqn:evidenceproduct}).
The first integral is estimated by a BIC type approximation denoted by $\textrm{BIC}_{lr}$ for logistic regression,
\begin{equation}\label{biclr}
  \log \left \{ \int_\beta \pi(\Y|\hat{\Z},\beta) \pi(\beta)d\beta\right\} \approx \frac{1}{2}\textrm{BIC}_{lr} = \frac{1}{2} \left(2\log\left\{\pi(\Y|\hat{\Z},\hat \beta(\hat\Z))\right\} - d_{lr} 
\log\left\{n_{lr}\right\}\right),
\end{equation}
 where  $\hat \beta(\hat\Z)$ is the maximum likelihood estimator of $\beta$ given a fixed posterior estimate of the latent locations $\hat \Z$,
$n_{lr}$ is the number of ties in the network
% or the number of possible ties $n\times (n-1)$ 
and $d_{lr}$ is the dimension of $\beta$, the number of parameters in the logistic regression model. 
Actor covariate data may be included in the logistic regression model (Handcock \textit{et al} \citeyear{handcocketal07}), in which case $d_{lr}>1$.
The second integral in equation (\ref{eqn:evidenceproduct}) is approximated using a similar BIC approximation denoted  $\textrm{BIC}_{lp}$ for
 the latent positions,
\begin{equation}\label{biclp}
 \log \left \{\int_{\bftheta} \pi(\hat{\Z}|\bftheta) \pi(\bftheta) d\bftheta \right\}\approx \frac{1}{2} \textrm{BIC}_{lp} =  \frac{1}{2} \left(2\log\{\pi(\hat{\Z}|\hat \bftheta(\hat\Z))\} - d_{lp} \log\{n\})\right),
\end{equation}
 where $\hat \bftheta(\hat\Z)$ is the maximum likelihood estimator of $\bftheta$ given the fixed posterior estimate of the latent positions $\hat \Z$ 
and $d_{lp}$ is the number of parameters in the mixture model.

A similar approach is carried out in the variational Bayesian framework \cite{salter:murphy12} where $\hat\Z$ is the modal variational posterior estimate of the
 latent positions.
This paper avoids the approximations of  equations (\ref{eqn:evidenceproduct}), (\ref{biclr}) and (\ref{biclp}) by modelling jointly, 
the number of components $G$ in a fully probabilistic Bayesian approach 
as well as exploring uncertainty in $\Z$ and $\beta$ using MCMC methods.
% \ul{and numerically integrating over all parameters including $\Z$ and $\beta$}
%  \ul{using MCMC methods.}
This is made computationally feasible by collapsing or integrating out the clustering  parameters from the model analytically (Section \ref{collapsing}).
Thus the marginal probability $\pi(G|\Y)$ can be estimated directly via MCMC sampling of the collapsed posterior.
%  details the collapsing or integrating out of parameters from the model to make this computationally feasible.
% Another issue with the approximations resulting from equations (\ref{biclr}) and (\ref{biclp}) is that they do not involve prior distributions, whereas
%  the model evidence (equation \ref{eqn:evidence}) depends on the model priors.  It is always the case 
% with BIC that it doesn't involved terms from the prior, so this sentence is redundant. 
% which can lead to further differences between the results of our methodology compared to current standard methods. 

\section{Collapsing the Model}\label{collapsing}
It is possible to integrate out or collapse the clustering parameters $\bftheta$ from the posterior distribution analytically by using the conjugate priors described in Section 
\ref{lpcm}. This yields a collapsed posterior distribution for the latent position cluster model,
\begin{eqnarray}\label{eqn:clpcm}
\pi(\Z,  \beta, \K, G|\Y)   &\propto &\int_{\bflambda}{\int_{\bfsigma}{\int_{\bfmu} {L(\Y|\Z,\beta)  \pi(\Z|\bfmu, \bfsigma^2,\K,G) }}} 
\pi(\bfmu|\bfsigma^2,G)\pi(\bfsigma^2|G)\pi(\K | \bflambda,G) \nonumber\\
    & &  \;\;\;\; \times \pi(\bflambda|G) \pi(\beta) \pi(G) \, \mbox{d}\bfmu \, \mbox{d}\bfsigma \, \mbox{d}\bflambda \nonumber\\
&=& L(\Y|\Z,\beta)  \pi(\Z|\K,G)  \pi(\K |G) \pi(\beta) \pi(G) \nonumber\\
&=&
 \displaystyle\prod_{i=1}^n{} \displaystyle\prod_{j \neq i} \frac{\exp\left\{y_{ij}\left(\beta-||\z_i - \z_j||\right)\right\}}{1+\exp\left\{\beta-||\z_i - \z_j||\right\}} \nonumber\\
&&\times  \prod_{g=1}^G\left(\frac{\Gamma\left(\frac{n_gd+\alpha}{2}\right)}{ \left(n_g+\frac{1}{\omega^2}\right)^{\frac{d}{2}}}\left(\delta+\sum_{i:k_i=g} \|z_i\|^2 - 
\frac{\|\sum_{i:k_i=g} z_i\|^2}{\left(n_g+\frac{1}{\omega^2}\right)}\right)^{-\left(\frac{n_gd+\alpha}{2}\right)}\right) \nonumber\\
&&\times \frac{\Gamma(G\nu)}{\Gamma(\nu)^G} \pi^{-\frac{dn}{2}}\frac{\left(\delta\right)^{\frac{G\alpha}{2}} }{\Gamma\left(\frac{\alpha}{2}\right)^G}
(\omega^2)^{-\frac{Gd}{2}} \frac{\prod_{g=1}^G \Gamma(n_g+\nu)}{\Gamma(n+G\nu)} \nonumber\\
%
%pi(\beta) 
% \beta  \sim \mathrm{Normal}_p(\xi, \psi)
&&\times \frac{1}{\sqrt {2\pi\psi  } }\exp\left\{-\frac{(\beta-\xi)^2}{2\psi} \right\} 
%\pi(G)=Poiss(1)
\times \frac{\exp\{-1\}}{G!},
\end{eqnarray}
where $n_g=\sum_{i=1}^n \mathbbm{1}(k_i=g)$.  Full details of the integration is given in the Appendix. 

This is similar to the approach of Nobile and Fearnside \citeyear{nob:fearn07} and Wyse and Friel \citeyear{wyse:friel12}, where 
the allocation sampling algorithm was developed for the collapsed finite mixture model and latent block models, respectively. 
This paper extends the approach to the latent position cluster model for social networks,
where the latent actor locations are analogous to the observed data in Nobile and Fearnside \citeyear{nob:fearn07}.  
The collapsed posterior for the latent position cluster model (equation \ref{eqn:clpcm}) depends on $\Z$ and $\K$, the latent positions and the allocation vector, respectively. In particular, $\K$ is of fixed dimension, but crucially, it provides
information on the number of components in the model. In this way, it is possible to carry out  trans-model inference 
 over a fixed dimensional parameter space, unlike reversible jump MCMC \cite{richardson:green97} for the full model which involves algorithmic moves of variable dimension.
Additionally, the collapsed posterior involves a much reduced parameter space, since the component means, variance and mixing weights are analytically integrated out.
% The number of parameters to be estimated is $(d+1)n+2$ compared to $(d+1)n+2+G(d+2)$ for the full model. 
The advantage of our approach is 
improved computational efficiency, reduced parameter storage requirements and a reduction in variability due to the removal of the uncertainty associated with the clustering 
parameters which have been integrated out of the model.

\section{A trans-model algorithm for the collapsed latent position cluster model}\label{algs}

Markov chain Monte Carlo sampling of the collapsed posterior distribution for the latent position cluster model
 is carried out using a Metropolis-within-Gibbs algorithm.
As full-conditional distributions for the positions $\Z$ and intercept $\beta$ are not of standard form, Metropolis-Hastings-within-Gibbs updates are required.
A standard Gibbs update is carried out to update the allocation vector $\K$ and we suggest a further $3$ Metropolis-Hastings-within-Gibbs moves to update $\K$ without changing 
the number of components in the model. Finally, a trans-model ejection/absorption move proposes the addition or removal of a component, 
changing only the fixed dimensional allocation vector $\K$.
The moves are similar to the approach of Nobile and Fearnside \citeyear{nob:fearn07} and Wyse and Friel \citeyear{wyse:friel12}.

\subsection{Metropolis-Hastings-within-Gibbs update for the latent positions $\Z$}\label{zupdategibbs}
% \paragraph{Metropolis-Hastings update of latent positions $\Z$}:}

Metropolis-Hastings-within-Gibbs sampling is carried out using the collapsed full-conditional distribution for the positions $\Z$,
\begin{eqnarray*}
\pi(\Z|\K,\beta,G,\Y) &\propto&  L(\Y|\Z,\beta)\pi(\Z|\K,G) \\
&\propto&\displaystyle\prod_{i=1}^n{} \displaystyle\prod_{j \neq i} \frac{\exp\{y_{ij} \left(\beta-||\z_i - \z_j||\right)\}}{1+\exp\{\beta-||\z_i - \z_j||\}}\\
&&\times  \prod_{g=1}^G\left(\frac{\Gamma\left(\frac{n_gd+\alpha}{2}\right)}{ \left(n_g+\frac{1}{\omega^2}\right)^{\frac{d}{2}}}\left(\delta+\sum_{i:k_i=g} \|\z_i\|^2 - \frac{\|\sum_{i:k_i=g} \z_i\|^2}
{\left(n_g+\frac{1}{\omega^2}\right)}\right)^{-\left(\frac{n_gd+\alpha}{2}\right)}\right).\\
% &&\times \prod_{g=1}^G \frac{\Gamma(n_g+\nu)}{\Gamma(n+G\nu)}.\\
% &&\times \prod_{g=1}^G \Gamma(n_g+\nu).\\
\end{eqnarray*}
%   \item 
A full sweep consists of visiting each $\z_i$ for $i=1,\ldots,n$ and proposing an update of $\z_i$ which is accepted or rejected using the usual Metropolis-Hastings 
accept/reject
probability, as outlined in Update (\ref{alg:z}).
%\SetAlgorithmName{}{}{}??
% \begin{algorithm}
% %  \newalgname{Pseudo Algorithme}
% \nocaptionofalgo
% At iteration $t$,
% \For{$i=1,\dots, n$}{
% \begin{itemize}
%  \item  [-] Propose $\z_i' \sim q(\z_{i}^t \to \z_i')$, where the proposal distribution $q$ is a Multivariate \\ Normal distribution $f(\z_i' ;\z_{i}^t,\sigma_\Z^2\I_d)$,
% with mean $\z_i^t$ and covariance matrix $\sigma_\Z^2 \I_d$;
%  \item  [-] Accept  $\z_i^{t+1}=\z_i'$ with probability $\min (1 ,\alpha)$ where, 
% \begin{eqnarray*}
%  \alpha =
% % &=& \frac{\pi(new)}{\pi(old)} \frac{p(new \to old)}{p(old \to new)} \\
% \frac{\pi(\Z'|\K, \beta, G,\Y)}{\pi(\Z^t|\K, \beta, G,\Y)} \frac{q( \z_i' \to \z_{i}^t)}{q(\z_{i}^t \to \z_i')}.\\
% \end{eqnarray*}
% Otherwise set $\z_{i}^{t+1}=\z_{i}^t$.
% \end{itemize}
% }
% %   Update membership vector $\K^{(t)}$ using Move 1 (Section\ref{move2})\;
% %   Update membership vector $\K^{(t)}$ using Move 1 (Section\ref{move3})\;  
% %   Update membership vector $\K^{(0)}$ using an absorption/ejection move (Section\ref{absejmove})\;
% % }
% \caption{\textbf{Update 1:} Metropolis Hastings update of the actor locations $\Z$ \label{alg:z}}
% \end{algorithm}

 \begin{algorithm}[H]
\nocaptionofalgo
% \SetLine
% \KwData{this text}
% \KwResult{how to write algorithm with \LaTeX2e }
At iteration $t$\;
\For{$i=1,\dots, n$}{
Propose $\z_i' \sim q(\z_{i}^t \to \z_i')$, where the proposal distribution $q$ is a Multivariate  Normal distribution $f(\z_i' ;\z_{i}^t,\sigma_\Z^2\I_d)$,
with mean $\z_i^t$ and covariance matrix $\sigma_\Z^2 \I_d$\;
% \eIf{understand}{
 Accept  $\z_i^{t+1}=\z_i'$ with probability $\min (1 ,\alpha)$ where,
\begin{eqnarray*}
 \alpha =
\frac{\pi(\Z'|\K, \beta, G,\Y)}{\pi(\Z^t|\K, \beta, G,\Y)} \frac{q( \z_i' \to \z_{i}^t)}{q(\z_{i}^t \to \z_i')};\\
\end{eqnarray*}
% }{
Otherwise set $\z_{i}^{t+1}=\z_{i}^t$.
% }
}
\caption{\textbf{Update 1:} Metropolis Hastings update of the actor locations $\Z$ \label{alg:z}}
\end{algorithm}

\subsection{Metropolis-Hastings-within-Gibbs update for the intercept $\beta$}\label{betaupdate}
% \paragraph{Metropolis-Hastings update for intercept $\beta$:}
The collapsed full-conditional distribution for the intercept parameter $\beta$ is not of standard form,
\begin{eqnarray*}
\pi(\beta|\Z, \Y) &\propto& \pi(\Y|\Z,\beta)\pi(\beta) \\
&\propto& \displaystyle\prod_{i=1}^n{} \displaystyle\prod_{j \neq i} \left(\frac{\exp\{y_{ij} \left(\beta-||\z_i - \z_j||\right)\}}{1+\exp\{\beta-||\z_i - \z_j||\}}\right)\exp\left\{-\frac{(\beta-\xi)^2}{2\psi} \right\} .
\end{eqnarray*}
Thus the Metropolis-Hastings-within-Gibbs Update (\ref{alg:beta}) is executed.

% \begin{algorithm}[h]
% \nocaptionofalgo
% At iteration $t$,
% \begin{itemize}
%  \item  [-] Propose $\beta' \sim q(\beta^t \to \beta')$, where the proposal distribution $q$ is a Normal  
% distribution \\ $f(\beta' ;\beta^t,\sigma_\beta^2)$,
% with mean $\beta^t$ and variance $\sigma_\beta^2$;
%  \item  [-] Accept  $\beta^{t+1}=\beta'$ with probability $\min (1,\alpha)$ where, 
% \begin{eqnarray*}
%  \alpha =\frac{\pi(\beta'|\Z, \Y)} {\pi(\beta_t|\Z, \Y)}
% % &=& \frac{\pi(new)}{\pi(old)} \frac{p(new \to old)}{p(old \to new)} \\
%  \frac{q( \beta' \to\beta^t)}{q(\beta^t \to \beta')};\\
% \end{eqnarray*}
% Otherwise set $\beta^{t+1}=\beta^t$.
% \end{itemize}
% %   Update membership vector $\K^{(t)}$ using Move 1 (Section\ref{move2})\;
% %   Update membership vector $\K^{(t)}$ using Move 1 (Section\ref{move3})\;  
% %   Update membership vector $\K^{(0)}$ using an absorption/ejection move (Section\ref{absejmove})\;
% % }
% \caption{\textbf{Update 2:} Metropolis Hastings update of the intercept parameter $\beta$ \label{alg:beta}}
% \end{algorithm}

 \begin{algorithm}[H]
\nocaptionofalgo
% \SetLine
% \KwData{this text}
% \KwResult{how to write algorithm with \LaTeX2e }
At iteration $t$\;
Propose $\beta' \sim q(\beta^t \to \beta')$, where the proposal distribution $q$ is a Normal  
distribution $f(\beta' ;\beta^t,\sigma_\beta^2)$,
with mean $\beta^t$ and variance $\sigma_\beta^2$;\\% \eIf{understand}{
 Accept  $\beta^{t+1}=\beta'$ with probability $\min (1,\alpha)$ where, 
\begin{eqnarray*}
 \alpha =\frac{\pi(\beta'|\Z, \Y)} {\pi(\beta_t|\Z, \Y)}
% &=& \frac{\pi(new)}{\pi(old)} \frac{p(new \to old)}{p(old \to new)} \\
 \frac{q( \beta' \to\beta^t)}{q(\beta^t \to \beta')};
\end{eqnarray*}
 \\
Otherwise set $\beta^{t+1}=\beta^t$.
% }
\caption{\textbf{Update 2:} Metropolis Hastings update of the intercept parameter $\beta$ \label{alg:beta}}
\end{algorithm}

% A proposal distribution $q$ is used to sample $\beta_{t+1}$ via Metropolis-Hastings sampling, $\beta' \sim q(\beta^t, \beta')$,
% where $q$ is a normal distribution centred on $\beta^t$ with proposal variance 
% $\sigma_\beta^2$.
% With  probability  $\min (1 \wedge \alpha*)$  the proposed $\beta'$ is accepted at iteration 
% ${t+1}$. Otherwise remain at $\beta_{t+1}=\beta_{t}$ The Metropolis-Hastings probability is 
% \begin{eqnarray*}
% \label{betaupdate}
% % \alpha=   \frac{\pi(\beta'|\Z, \Y) q(\beta'|\beta_t)} {\pi(\beta_t|\Z, \Y)q(\beta_t|\beta')} symmetry cancels
% \alpha &=&   \frac{\pi(\beta'|\Z, \Y)} {\pi(\beta_t|\Z, \Y)}\\
% % \alpha=   \frac{\pi(\beta|\Z, \Y) q(\beta'|\xi, \psi)} {\pi(\beta|\Z, \Y)q(\beta_t|\xi, \psi)}.
% % &=&   \frac{\pi(\Y|\Z_{t+1},\beta') q(\beta'|\beta_t)} {\pi(\Y|\Z_{t+1},\beta_t)q(\beta_t|\beta')}\\
% &=& \displaystyle\prod_{i=1}^n{} \displaystyle\prod_{j \neq i} \frac{\exp\{y_{ij} \left(\beta'-||\z_i - \z_j||\right)\}}{1+\exp\{\beta'-||\z_i - \z_j||\}}\frac{1}{\sqrt {2\pi\psi  } }\exp\left\{-\frac{-(\beta'-\xi)^2}{2\psi} \right\} \\
% &\times&\left(\displaystyle\prod_{i=1}^n{} \displaystyle\prod_{j \neq i} \frac{\exp\{y_{ij} \left(\beta_t-||\z_i - \z_j||\right)\}}{1+\exp\{\beta_t-||\z_i - \z_j||\}}\frac{1}{\sqrt {2\pi\psi  } }\exp\left\{-\frac{-(\beta_t-\xi)^2}{2\psi} \right\} \right)^{-1}.
% \end{eqnarray*}.

\subsection{Moves to update the cluster membership vector $\K$}\label{kupdate}

\subsubsection{Gibbs update}
% \paragraph{$\K$ update using Gibbs}
To update the allocation vector $\K$, a standard Gibbs update is performed using its full-conditional distribution,
\begin{eqnarray*}
% \pi(\K |G) &=&\int_{\bflambda}\pi(\K | \bflambda,G)=\int_{\bflambda}\prod_{g=1}^G (\lambda_g^{n_g+\nu-1})\, \mbox{d}{\bflambda}=\frac{\prod_{g=1}^G \Gamma(n_g+\nu)}{\Gamma(\sum_g(n_g+\nu))}\\
\pi(\K |\Z,G,\phi) &\propto& \pi(\Z|\K,G,\phi)\pi(\K|G,\phi) \\
&\propto&
%  \times \displaystyle\prod_{j \neq i} \frac{\exp\{y_{ij} \left(\beta-||\z_i' - \z_j||\right)\}}{1+\exp\{\beta-||\z_i' - \z_j||\}}\\
\prod_{g=1}^G\left(\frac{\Gamma(\frac{n_gd+\alpha}{2})}{ (n_g+\frac{1}{\omega^2})^{\frac{d}{2}}}\left(\delta+\sum_{i:k_i=g} \|\z_i\|^2 - \frac{\|\sum_{i:k_i=g} \z_i\|^2}{(n_g+\frac{1}{\omega^2})}\right)^{-(\frac{n_gd+\alpha}{2})}\right)
% \frac{\prod_{g=1}^G \Gamma(n_g+\nu)}{\Gamma(N+G\nu)}.\\
\Gamma(n_g+\nu).\\
% && \times \int_{\bflambda}\pi(\K | \bflambda,G)=\int_{\bflambda}\prod_{g=1}^G (\lambda_g^{n_g+\nu-1})\, \mbox{d}{\bflambda}=\frac{\prod_{g=1}^G \Gamma(n_g+\nu)}{\Gamma(\sum_g(n_g+\nu))}
% =\frac{\prod_{g=1}^G \Gamma(n_g+\nu)}{\Gamma(N+G\nu))}\\.
\end{eqnarray*}                 
%  \begin{eqnarray*}
% % \pi(\K |G) &=&\int_{\bflambda}\pi(\K | \bflambda,G)=\int_{\bflambda}\prod_{g=1}^G (\lambda_g^{n_g+\nu-1})\, \mbox{d}{\bflambda}=\frac{\prod_{g=1}^G \Gamma(n_g+\nu)}{\Gamma(\sum_g(n_g+\nu))}\\
% \pi(\K |G) &=&\int_{\bflambda}\pi(\K | \bflambda,G) \mbox{d}{\bflambda}\\
% &=& \frac{\prod_{g=1}^G \Gamma(n_g+\nu)}{\Gamma(\sum_g(n_g+\nu))}\\
% && \times \displaystyle\prod_{j \neq i} \frac{\exp\{y_{ij} \left(\beta-||\z_i' - \z_j||\right)\}}{1+\exp\{\beta-||\z_i' - \z_j||\}}\\
% &&\times  \prod_{g=1}^G\left(\frac{\Gamma(\frac{n_gd+\alpha}{2})}{ (n_g+\frac{1}{\omega^2})^{\frac{d}{2}}}\left(\delta+\sum_{i:k_i=g} \|\z_i\|^2 - \frac{\|\sum_{i:k_i=g} \z_i\|^2}{(n_g+\frac{1}{\omega^2})}\right)^{-(\frac{n_gd+\alpha}{2})}\right)\\
% % && \times \int_{\bflambda}\pi(\K | \bflambda,G)=\int_{\bflambda}\prod_{g=1}^G (\lambda_g^{n_g+\nu-1})\, \mbox{d}{\bflambda}=\frac{\prod_{g=1}^G \Gamma(n_g+\nu)}{\Gamma(\sum_g(n_g+\nu))}
% % =\frac{\prod_{g=1}^G \Gamma(n_g+\nu)}{\Gamma(N+G\nu))}\\.
% \end{eqnarray*}
A full sweep consists of visiting each $k_i$ for $i=1,\ldots,n$ and carrying out Update \ref{alg:k}.

\begin{algorithm}
\nocaptionofalgo
At iteration $t$\;
\For{$i=1,\dots, n$}{
 Compute $\pi_g(\K^* |\Z,G,\phi)$ where $\K^*=(k_1^{(t)},\ldots,k_i=g,\ldots,k_n^{(t-1)})$ for $g=1,\dots, G$\;
 Sample $k_i^{(t)}$ from the vector of weights $\frac{\pi_g(\K^* |\Z,G,\phi)}{\sum_{g=i}^G \pi_g(\K^* |\Z,G,\phi)}$ for $g=1,\dots, G$.
% Compute $\pi(k_i=j|g,k_{-i},\phi) = \frac{f_j}{\sum_j=i^G f_j}$ for $j=1,\dots, G$

}
%   Update membership vector $\K^{(t)}$ using Move 1 (Section\ref{move2})\;
%   Update membership vector $\K^{(t)}$ using Move 1 (Section\ref{move3})\;  
%   Update membership vector $\K^{(0)}$ using an absorption/ejection move (Section\ref{absejmove})\;
% }
\caption{\textbf{Update 3:} Gibbs sampling of the actor allocation vector $\K$ \label{alg:k}}
\end{algorithm}

A further $3$ moves are proposed to update $\K$, without changing the number of components in the model. These mimic the allocation sampling algorithm of Nobile and 
Fearnside \citeyear{nob:fearn07} and serve to update several actor allocations simultaneously, searching more easily across the discrete set of possible allocation vectors.

\subsubsection{Move 1}\label{move1}
% \paragraph{Move 1:}
The first Metropolis-Hastings move to update $\K$ without changing the number of groups reallocates 
the actors of two components $j_1$ and $j_2$ selected at random from the $G$ available groups. 
Observations in both groups are re-allocated to component $j_1$ with
probability $p$ and to component $j_2$ with probability $1-p$ where $p$ is $\textrm{Beta}(1,1)$ distributed. 
The current and proposed allocation vectors are $\K$ and $\K'$ respectively and the proposal is symmetric.
The move is accepted with probability $\min(1,\alpha)$ where,
\begin{displaymath}
 \alpha=\frac{\pi(\Z|G,\K',\phi)}{\pi(\Z|G,\K,\phi)}.
\end{displaymath}

% \begin{algorithm}[h]
% At iteration $t$,
% \For{$i=1,\dots, n$}{
% \begin{itemize}
%  \item [-] Randomly select $j_1$ and $j_2$ among the $g$ available groups. 
%  \item [-] Draw $p \sim \textrm{beta}(1,1)$. 
%  \item  [-] Propose $\K'$, where if actor $i \in j_1$ or $i \in j_2$, set $k_i=j_1$ with probability $p$ or $k_i=j_2$ with probability $1-p$.
%  \item  [-] Accept  $\K^{t+1}=\K'$ with probability $\min (1 \wedge \alpha)$ where, 
% \begin{eqnarray*}
%  \alpha = \alpha=\frac{\pi(\Z|G,\K',\phi)}{\pi(\Z|G,\K,\phi)}.
% \end{eqnarray*}
% Otherwise set $\K^{t+1}=\K^t$.
% \end{itemize}
% %   Update membership vector $\K^{(t)}$ using Move 1 (Section\ref{move2})\;
% %   Update membership vector $\K^{(t)}$ using Move 1 (Section\ref{move3})\;  
% %   Update membership vector $\K^{(0)}$ using an absorption/ejection move (Section\ref{absejmove})\;
% }
% \caption{Metropolis Hastings update of allocation vector $\K$ (Move 1) \label{alg:m1}}
% \end{algorithm}

\subsubsection{Move 2}\label{move2}
% \paragraph{Move 2:}

The second Metropolis-Hastings update proposes to move a subset of members of one component at random to another component.
 The idea is that, if these observations are already grouped together into one component, 
then they may be similar in nature. Thus it may be possible to move them together at the same time to another component.
Components $j_1$ and $j_2$ are randomly selected among the $G$ available groups. 
If $n_{j_1}\neq 0$, $m$ random observations are selected from component $j_1$ and proposed to move to component $j_2$, 
where $m$ is drawn from a Uniform distribution on $\{1,\ldots,n_{j_1}\}$.
 The proposal is accepted with probability $\min(1,\alpha)$ where,
\begin{displaymath}
 \alpha=\frac{\pi(\Z|G,\K',\phi)}{\pi(\Z|G,\K,\phi)} \frac{q(\K'\to \K)}{q(\K \to \K')}
\end{displaymath}
and where,
\begin{displaymath}
 \frac{q(\K'\to \K)}{q(\K \to \K')}=\frac{n_{j_1}}{n_{j_2}+m}\frac{n_{j_1}!n_{j_2}!}{(n_{j_1}-m)!(n_{j_2}+m)!}.
\end{displaymath}

\subsubsection{Move 3}\label{move3}
% \paragraph{Move 3:}

The third proposal is similar to the first. Again, the actors of randomly selected components $j_1$ and $j_2$ are allocated
to one of the two groups. However, the probability $p$ is no longer constant for all actors. 
Instead, in a random sequence, actor $i$ is proposed to move with probability $p_j^{(i)}$ for $j \in \{j_1,j_2\}$, where $p_j^{(i)}$  is proportional to the probability that component 
$j$ generated the $i$-th observation, conditional on its value $\z_i$, on the previously re-allocated observations and on their new allocations. 
See Appendix $A2$ of Nobile and Fearnside \citeyear{nob:fearn07} for further details. 
 The proposal is accepted with probability $\min(1,\alpha)$ where,
\begin{displaymath}
 \alpha=\frac{\pi(\Z|G,\K',\phi)}{\pi(\Z|G,\K,\phi)} \frac{q(\K'\to \K)}{q(\K \to \K')}
\end{displaymath}
and where,
\begin{displaymath}
 \frac{q(\K'\to \K)}{q(\K \to \K')}=\prod_{i}\frac{p_{k_i}^{(i)}}{p_{k'_i}^{(i)}}.
\end{displaymath}

\subsubsection{Absorption/ejection moves}\label{absejmove}
% \paragraph{Absorption/Ejection move:}
This pair of trans-model Metropolis-Hastings-within-Gibbs moves
involve the addition or removal of a component.
To add a component to the model, a new component $j'_2$ is ejected from a randomly selected existing component $j_1$.
The members of component $j_1$ are allocated to component $j'_1$ with probability $p$ and to component $j'_2$ with probability $1-p$, where $p$ is Beta($a,a$) distributed. 
In the reverse move, one component absorbs another. 
The absorption/ejection moves do not change the dimension of the parameter space since $\K$ is of fixed dimension. 

Suppose an absorption or ejection move is attempted from current state $\{\K,G\}$ to $\{\K',G'\}$, where 
where $G$ and $G'$ are the current and proposed number of groups in the model respectively.
The probability of choosing a split move is $p^e=(1,0.5,\ldots,0.5,0)$, where $p_1^e=1$ since a $1$ component model must be split and $p_{G_{\mathrm{max}}}^e = 0$ as it is not possible to
 split a $G_{\mathrm{max}}$ component model, where $G_{\mathrm{max}}$ is the maximum number of components allowed. This is a user specified value
taken to be $\lfloor n/2 \rfloor$ in our examples.

The move is accepted with probability  $\textrm{min} (1,\alpha)$ where,
% \begin{displaymath}
% \alpha= \frac{\pi(\K',G',\Z|\phi)}{\pi(\K,G,\Z|\phi)}   \frac{q(\{\K',G'\} \to \{\K',G'\})}{q(\{\K,G\} \to \{\K',G'\})},
% \end{displaymath}
\begin{displaymath}
\alpha= \frac{\pi(\Z'|\K',G',\phi) \pi(\K'|G',\phi) \pi(G')}{\pi(\Z|\K,G,\phi) \pi(\K|G,\phi) \pi(G) }
\frac{q(\{\K',G'\} \to \{\K,G\})}{q(\{\K,G\} \to \{\K',G'\})},
\end{displaymath}
% and where,
% \begin{displaymath}
%  \frac{q(\K'\to \K)}{q(\K \to \K')}=\prod_{i}\frac{p_{k_i}^{(i)}}{p_{k'_i}^{(i)}}.
% \end{displaymath}
% and where,
% \begin{displaymath}
% \pi(\K',G',\Z|\phi) = \pi(\Z|\K,G,\phi) \pi(\K|G,\phi) \pi(G) ,
% \end{displaymath}
and where,
\begin{displaymath}
\frac{ q(\{\K',G'\} \to \{\K,G\})}{q(\{\K,G\} \to \{\K',G'\})}=\frac{1-p_{G+1}^e}{p_G^e}\frac{2\Gamma(a)}{\Gamma(2a)}\frac{\Gamma(2a+n_{j1})}{\Gamma(a+n_{j'_1})\Gamma(a+n_{j'_2})}.
\end{displaymath}
The reverse absorb move is analogous with the probability of proposing an absorb move, $p^a=(0,0.5,\ldots,0.5,1)$ and acceptance probability $\textrm{min} (1,\alpha^{-1})$.
% See appendix $A2$ of Nobile \& Fearnside \citeyear{nob:fearn07} for further details.

\subsection{The collapsed sampling algorithm}
To sample from the collapsed latent position cluster model, we implement Algorithm (\ref{alg:clpcm}).
The output of the Markov chain at iteration $t$ is denoted by $(\Z^{(t)},\beta^{(t)},\K^{(t)})$.

\begin{algorithm}[h]
Initialise $(\Z^{(0)}, \beta^{(0)}, \K^{(0)})$\;
\For{$t=1,\dots, T$}{
  Update the latent actor locations $\Z^{(t)}$ using the Metropolis-Hastings-within-Gibbs Update (\ref{alg:z}) described in Section (\ref{zupdategibbs})\;
  Update the intercept $\beta^{(t)}$ using a Metropolis-Hastings-within-Gibbs Update (\ref{alg:beta}) described in Section (\ref{betaupdate})\;
  Update the cluster membership vector $\K^{(t)}$ without changing the number of components using a full sweep of 

\begin{itemize}
 \item [-] the Gibbs update described in Section (\ref{kupdate})\;
 \item [-] the Metropolis-Hastings Move 1 described in Section (\ref{move1})\;
\item [-] the Metropolis-Hastings Move 2 described in Section (\ref{move2})\;
\item [-] the Metropolis-Hastings Move 3 described in Section (\ref{move3})\;
\end{itemize}
 
Update the cluster membership vector $\K^{(t)}$ while simultaneously changing the number of components using 
\begin{itemize}
\item [-] the Absorption / ejection move described in Section (\ref{absejmove}).
\end{itemize}
%   Update membership vector $\K^{(t)}$ using Move 1 (Section\ref{move2})\;
%   Update membership vector $\K^{(t)}$ using Move 1 (Section\ref{move3})\;  
%   Update membership vector $\K^{(0)}$ using an absorption/ejection move (Section\ref{absejmove})\;
}
\caption{Sampling from the collapsed latent position cluster model\label{alg:clpcm}}
\end{algorithm}

\subsection{Post-processing}
Post processing is required due to the invariance of the likelihood to reflections, rotations and translations of the latent space and to the re-labelling of clusters. 
This is due to the fact that the actor positions appear in the likelihood as function of Euclidean distance only. 
% A maximum a posteriori (MAP) EXPLAIN estimate is used  as 
To address this problem, a Procrustes transformation \cite{sibson1978} is used to match each iteration to a reference configuration. 
The realisation of the Markov chain with the highest likelihood value is used as a reference configuration.
The likelihood invariance to the switching of cluster labels is corrected by iteratively minimising the cost associated with all possible label 
permutations using the square assignment algorithm of
Carpeneto and Toth \citeyear{carpaneto1980algorithm} as in Nobile and Fearnside \citeyear{nob:fearn07} 
and Wyse and Friel \citeyear{wyse:friel12}. 

% The allocation vectors $\K_t$ for iterations $t=1,\ldots T$ of the chain are ordered by increasing number of non-empty components.
% The cost matrix $C$ is calculated where,
% \begin{displaymath}
%  C(\K_1,\K_2) = \sum_{t=1}^{T-1}\sum_{i=1}{n}\I{k_i=k_i'}
% \end{displaymath}
% The code of Carpeneto and Toth \citeyear{carpaneto1980algorithm} uses a square assignment algorithm to minimize the total cost $C$ and returns the corresponding 
% permutation of allocation vectors. 
% T −1 n
% (t)
% (T )
% I zi = k1 , zi
% C(k1 , k2) =
% = k2 .
% t=1 i=1
% Then the more z(T ) disagrees with the vectors already processed, the higher this cost will
% be (see Appendix C). The square assignment algorithm of Carpaneto & Toth (1980) returns
% KT −1
% the permutation σ(·) of the labels in z(T ) which minimizes the total cost k=1 C(k
% 

\section{Results}\label{res}

The methods are illustrated using some well known social networks, Sampson's $18$ node network \cite{sampson68},  Zachary's $34$ node karate club network 
\cite{zachary77} and a $62$ node network of New Zealand Dolphins \shortcite{lusseauetal03}. The examples serve to illustrate our methodology, to highlight the 
importance of model uncertainty for well known social networks and to make comparisons with inference using \texttt{latentnet} \cite{Kriv:Hand07,Kriv:Hand13}
 and the variational approximation to the posterior using \texttt{VBLPCM}. 
% The use of \texttt{latentnet} is not feasible for the larger of our protein-protein interaction datasets of size $200$ but comparisons
% can be made to the variational Bayesian approximation \cite{salter:murphy12}.

Inference using \texttt{latentnet} involves sampling  from the full posterior of Handcock \textit{et al} \citeyear{handcocketal07}. The number of clusters $G$ 
 is fixed and inference is carried out separately for $G=1,\ldots,G_{\mathrm{max}}$. The BIC approximation to the model evidence is used to choose the `best' model.

The Variational Bayes approach to inference is implemented using \texttt{VBLPCM} \cite{salter:murphy12}.
The Kullback-Leibler divergence from an approximate fully factorised variational posterior to the true posterior distribution is minimized.
 Inference is carried out separately for $G=1,\ldots,G_{\mathrm{max}}$ component models.
 A good initialisation of the variational parameters is important \cite{salter:murphy12}.
The Fruchterman-Reingold layout is used to initialise the latent positions, followed by the use of \texttt{mclust} \cite{fraley:raftery02,fraley:raftery03}
to initialise the clustering parameters. 
% As Fruchterman-Reingold is an iterative process which is itself initialised using random a configuration, 
% different results will be found each time.
The Fruchterman-Reingold layout
algorithm is itself initialised using a random configuration, thus different results will be found each time.
 The variational approximation which is `closest' to the true posterior in terms of Kullback-Leibler divergence is chosen as the best $G$ component model 
from $10$ different initialisations. The BIC approximation to the model evidence is then used to choose the number of components (Section \ref{bicsec}).

\subsection{Sampson's monks}\label{resmonks}

Sampson \citeyear{sampson68} conducted a social science study of  $18$ monks in a monastery during the time of Vatican II. During the study, a political `crisis in 
the cloister' resulted in the expulsion of four monks and the voluntary departure of several others.
A directed network was recorded where each monk was asked to rank $3$ friends across $3$ points in time. For the purposes of illustrating our methodology,
 we use the aggregated version of this network widely used in social network analysis literature. However the extension of the latent position cluster model to
temporal networks is an open problem.

\begin{figure}[htp]
 \begin{center}
\includegraphics[width=7cm]{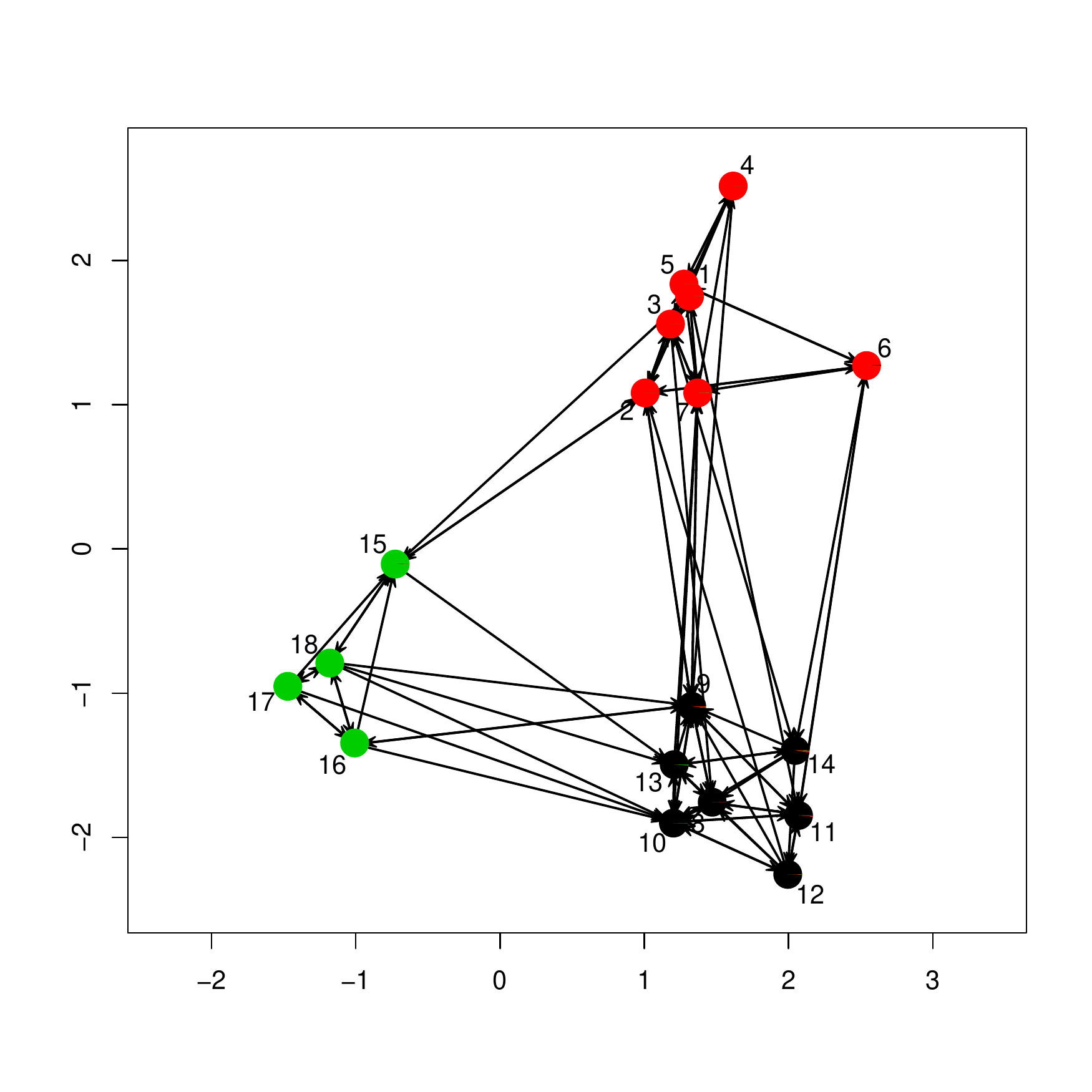}
\includegraphics[width=7cm]{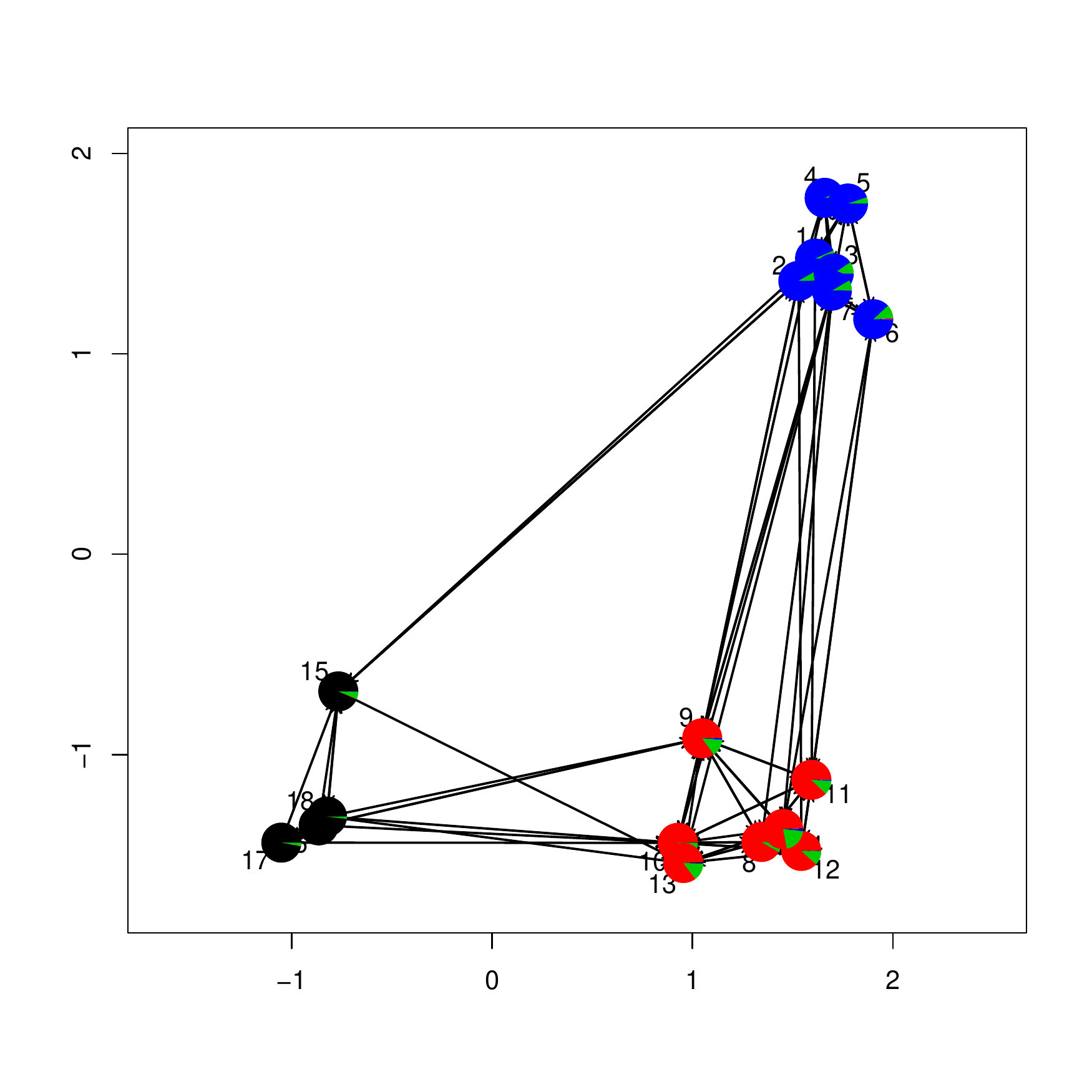}
 \end{center}
 \caption{Sampson's monks posterior mean actor positions using the collapsed sampler for the most probable $3$ and $4$ group models (left and right hand plots respectively) with a pie chart depicting uncertainty of cluster memberships.}% and latent positions using latent net for 2 group model}
\label{zmeanpiemonks}
\end{figure}

A $3$ or $4$ component model is widely accepted as the most suitable clustering for this data. 
The collapsed model inference was in agreement, with probabilities $0.79$ and $0.16$ for the $3$ and $4$ component models respectively. 
The sampling of the collapsed posterior as outlined in Algorithm (\ref{alg:clpcm}) took less than $1$ minute for $100,000$ draws, thinned by $10$.
The resulting posterior mean actor positions are shown in Figure \ref{zmeanpiemonks}, where arrows represent directed ties between nodes and pie charts indicate
the uncertainty in the cluster membership of actors. 

\begin{figure}
        \centering
        \begin{subfigure}[b]{0.3\textwidth}
                \centering
                \includegraphics[width=\textwidth]{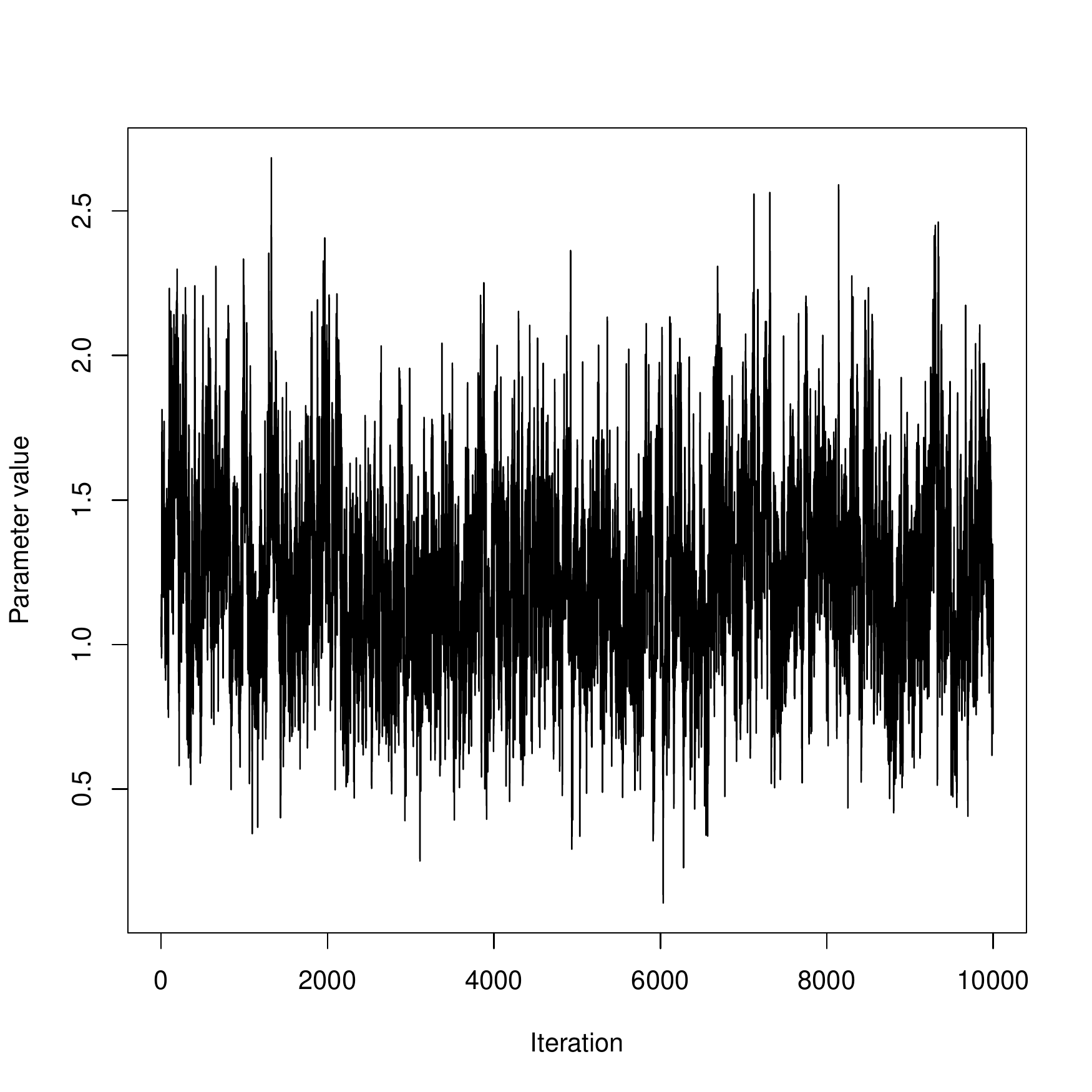}
                \caption{}\label{btr}
        \end{subfigure}%
%         \begin{subfigure}[b]{0.3\textwidth}
%                 \centering
%                 \includegraphics[width=\textwidth]{Monks15apr/gtrace.pdf}
%                 \caption{}\label{gtr}
%         \end{subfigure}%
        \begin{subfigure}[b]{0.3\textwidth}
                \centering
                \includegraphics[width=\textwidth]{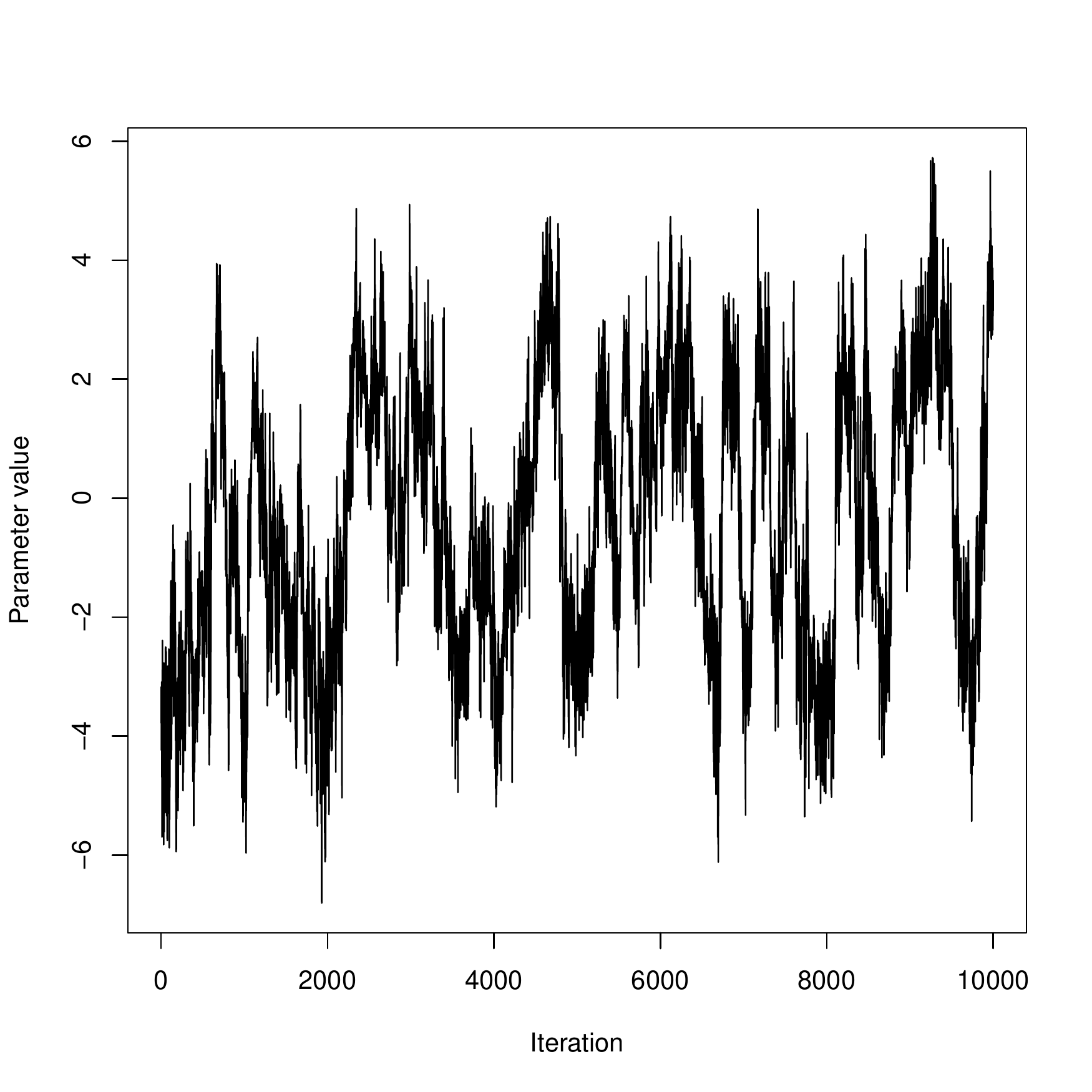}
                \caption{}
        \end{subfigure}
        \begin{subfigure}[b]{0.3\textwidth}
                \centering
                \includegraphics[width=\textwidth]{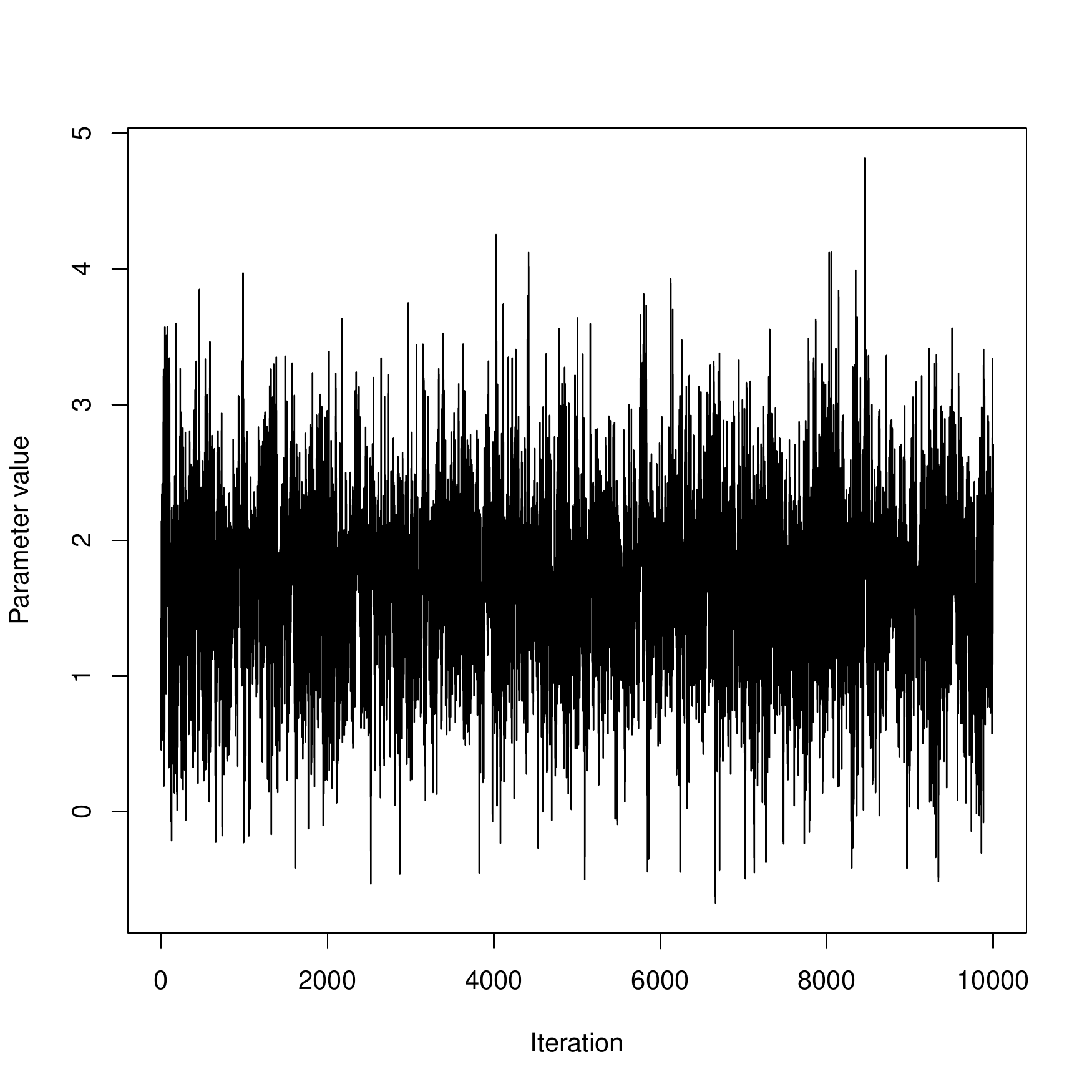}
                \caption{}
         \end{subfigure}
\caption{Trace plots of intercept $\beta$ (plot (a)) and one sample latent position $z_{(1,1)}$ pre- and post-Procrustes matching (plots (b), (c) respectively) for Sampson's monks.}
\label{tracemonks}
\end{figure}

 Proposal variances for the Metropolis-Hastings moves 
were $\sigma_z^2=0.7$ for the latent actor positions and $\sigma^2_\beta=0.5$ for the intercept.
Acceptance rates are displayed in Table \ref{acceptmonks}. 
Figure \ref{tracemonks} displays trace plots of the intercept $\beta$ and a sample actor position
 pre- and post-Procrustes matching demonstrating the effect of post processing the positions.
The choice of hyperparameters, $\delta=0.103$, $\alpha=2$ and $\nu=3$, are in line with Handcock \textit{et al} \citeyear{handcocketal07}.
A Poisson($1$) prior on the number of groups was used as per Nobile and Fearnside \citeyear{nob:fearn07} and the prior variance of the cluster means were 
dependent on $\sigma_g^2$ and scaled by a factor of $\omega^2=10$. The analysis was reasonably insensitive to the choice of hyperparameters since
the $3$ cluster model chosen for a wide range of values of $\phi$.
As described in Section \ref{lpcm}, the priors are not identical for each model which must be considered when 
comparing inference using the collapsed sampling, inference using \texttt{latentnet} and the variational approach using \texttt{VBLCPM}.

\begin{figure}
 \centering
\includegraphics[width=7cm]{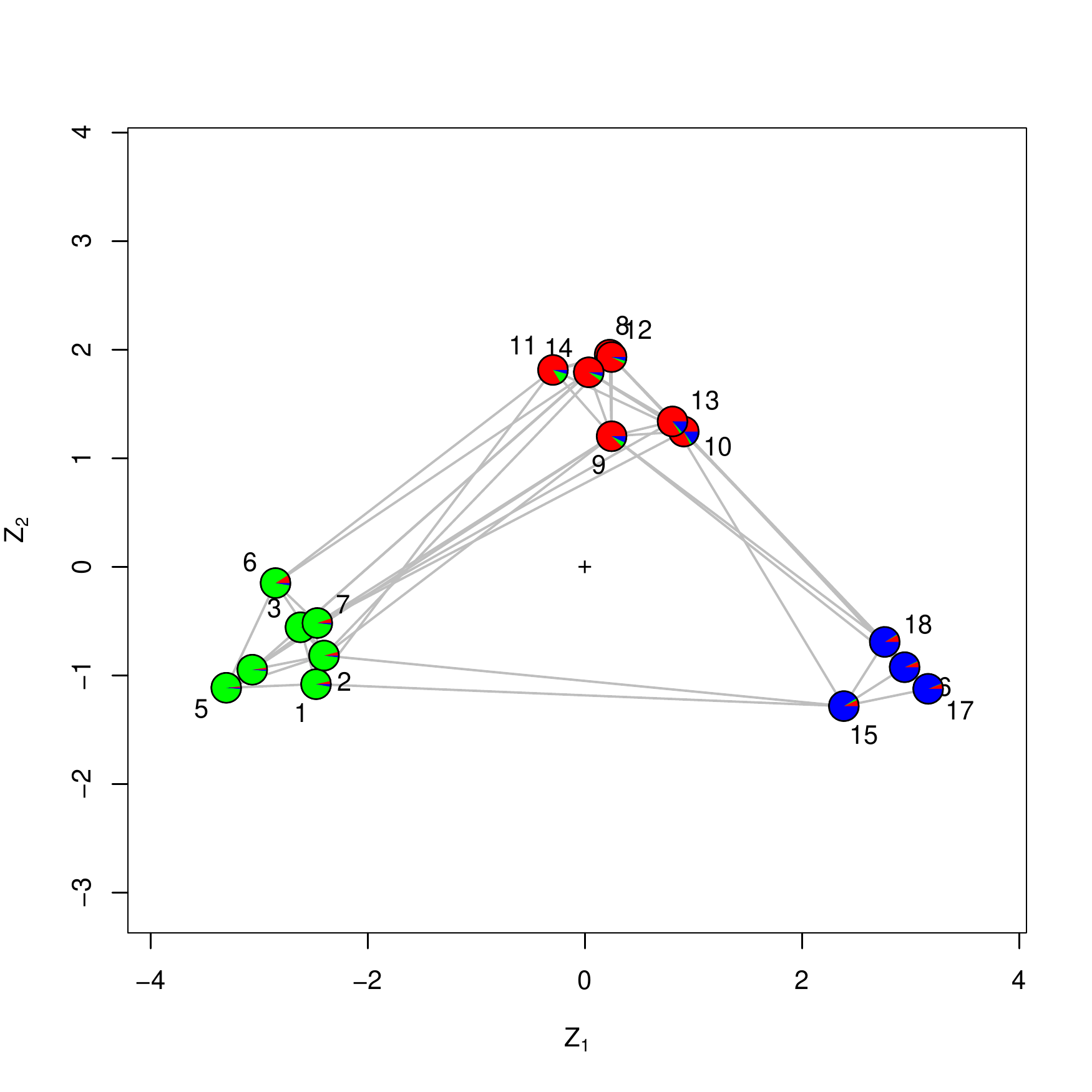}
\includegraphics[width=7cm]{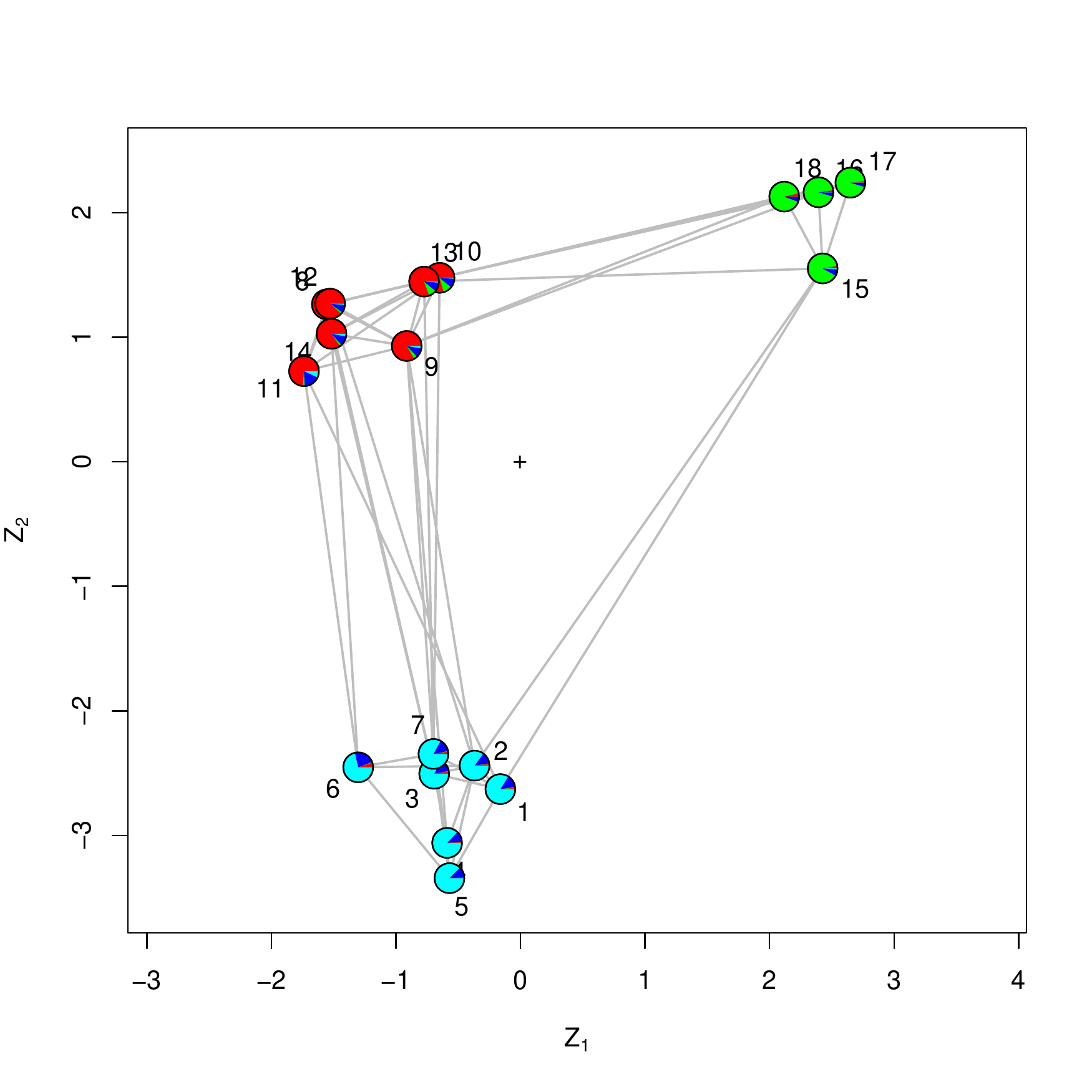}
\caption{Posterior mean latent positions for a $3$ and $4$ component model for Sampson's monks using \texttt{latentnet}.}
\label{lnetmonks}
\end{figure}

\begin{figure}
 \centering
 \includegraphics[width=7cm]{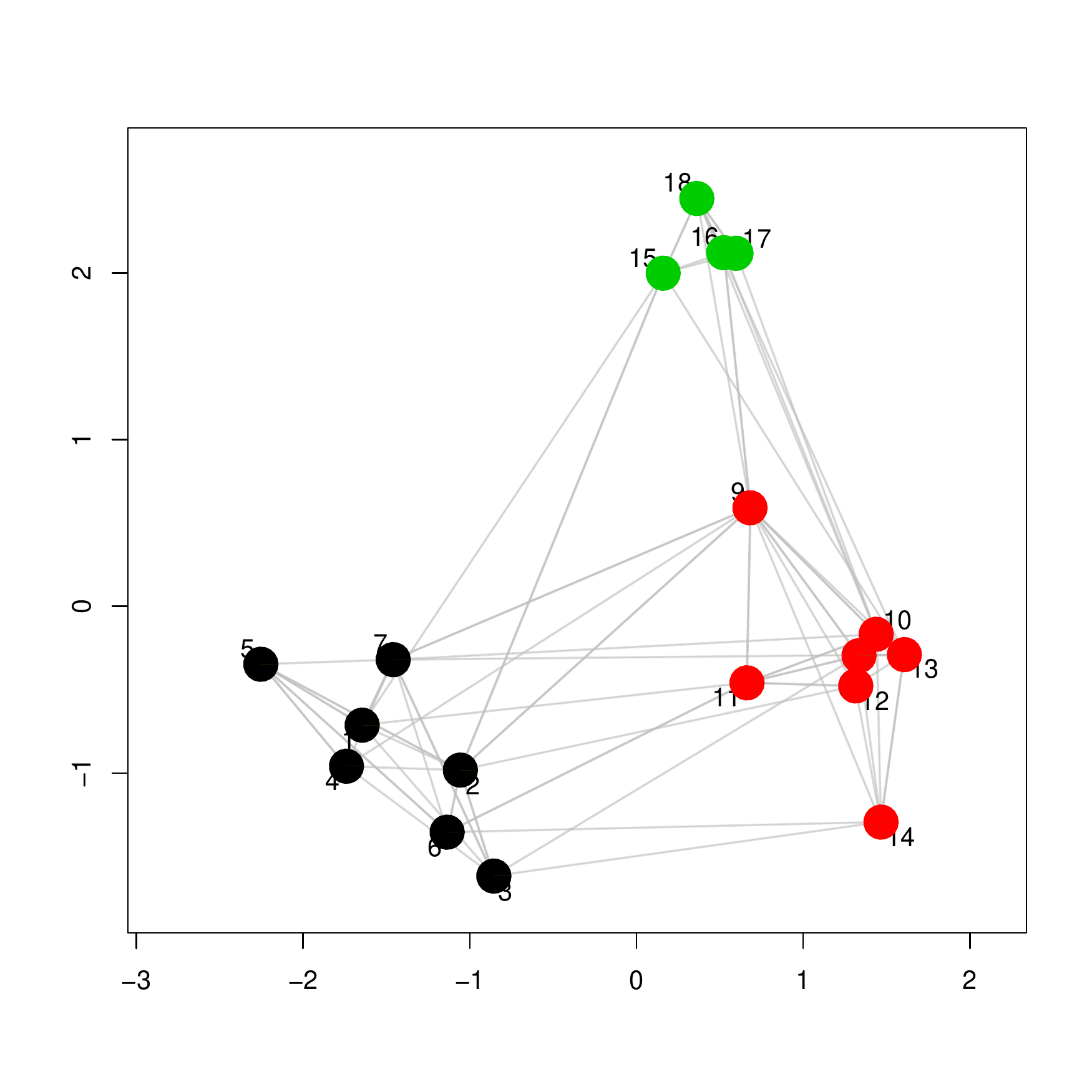} 
 \includegraphics[width=7cm]{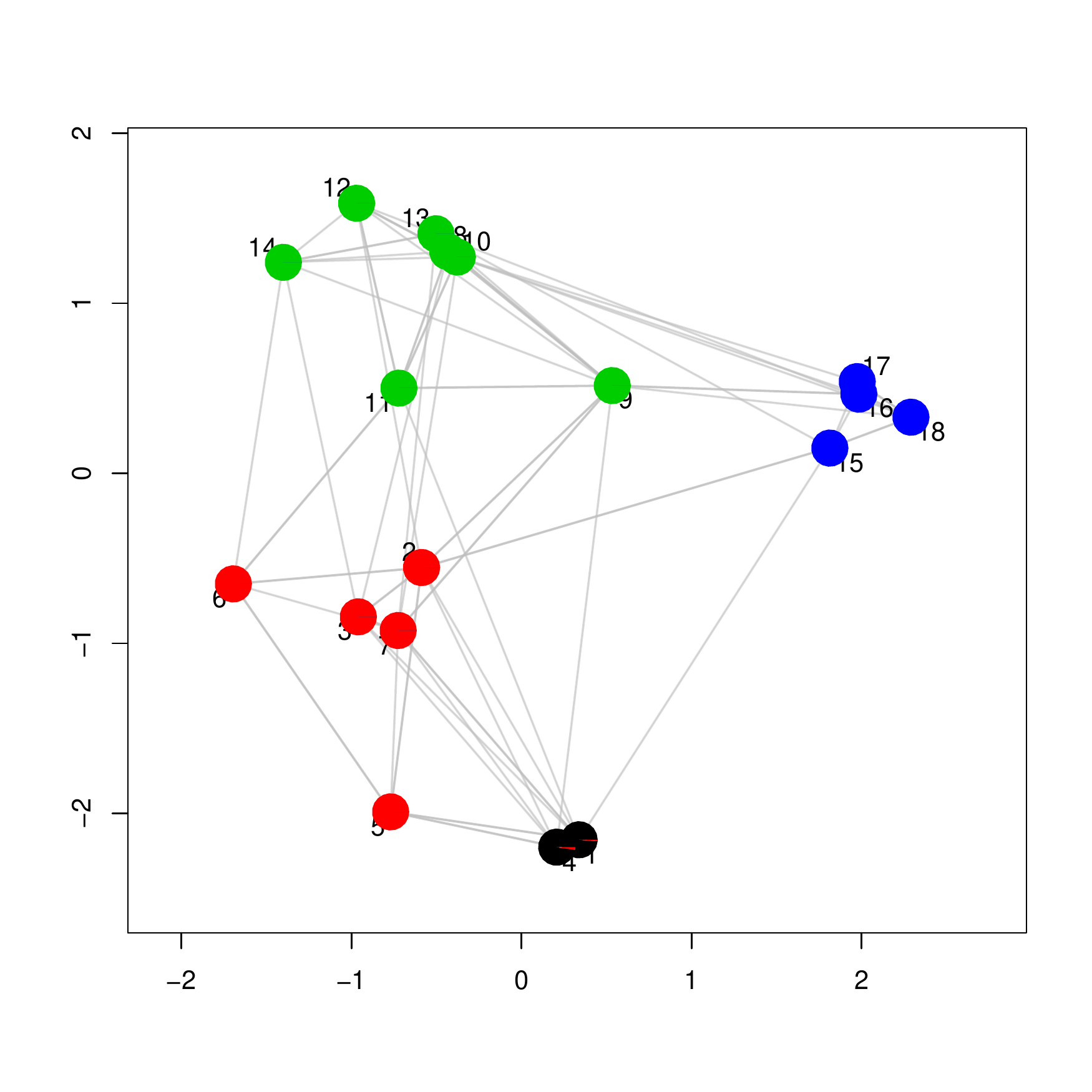} 
\caption{Variational Bayesian estimates of Latent positions for a $3$ and $4$ component model for Sampson's monks.}
\label{vbmonks}
\end{figure}

Posterior model probabilities for the collapsed method are displayed in Table \ref{pallmonks} together with approximate 
BIC values inferred using \texttt{latentnet} and \texttt{VBLPCM}. 
The lowest BIC value is the `best' model. All approaches favoured $3$ clusters in the network.
Sampling the full posterior using \texttt{latentnet} took $24$ minutes for $100,000$ 
draws of the $1$ to $5$ component models compared to $1$ minute for the collapsed sampling.  There is further agreement in that
the second lowest BIC values are for the $4$ component model. 
%For a fair comparison of
%timing it might be best to compare $500,000$ draws of the allocation sampler to this run since it is the likelihood calculation which is computationally intensive and 
%occurs at each iteration. This is still a $5$ fold improvement on computation time! 
The best estimate for the actor latent positions using \texttt{latentnet} for the $3$ and $4$ group models are displayed in Figure \ref{lnetmonks}. 
The variational approach took $50$ seconds to find the best of $10$ modal variational estimates using \texttt{VBLPCM} \cite{salter:murphy12}.
The $3$ and $4$ cluster variational fits are displayed in Figure \ref{vbmonks}.

% \begin{figure}[htp]
%  \begin{center}
% \includegraphics[width=7cm]{Monks15apr/zmeanpiecltied3g.pdf}
% \includegraphics[width=7cm]{Monks15apr/zmeanpiecltied4g.pdf}
%  \end{center}
%  \caption{Sampson's monks posterior mean actor positions using the collapsed sampler for the most probable $3$ and $4$ group models (left and right hand plots respectively) with a pie chart depicting uncertainty of cluster memberships.}% and latent positions using latent net for 2 group model}
% \label{zmeanpiemonks}
% \end{figure}

\begin{table}[htp]
\begin{center}
\begin{tabular}{| l|c|c|c|c|c|}
\hline
 & $G=1$ & $G=2$ & $G=3$ & $G=4$ & $G=5$ \\
\hline
 Collapsed model probabilities $\pi(G|\Y)$& 0.0005&0.0092&\underline{0.7886}&0.1604&0.1727 \\ 
  \texttt{latentnet} BIC&482.56&476.20&\underline{439.75}&443.51&447.19\\ 
 \texttt{VBLPCM} BIC&530.54& 497.51& \underline{473.55}& 486.49& 494.40\\
%   \texttt{VBLPCM} BIC& 541.69& 509.65&  \underline{484.71}& 502.41& 546.09\\
 \hline
\end{tabular}
\caption{Posterior model probabilities for Sampson's monks for the collapsed sampling and BIC values fitting $5$ models separately using \texttt{latentnet} and \texttt{VBLPCM}.
(The model underlined denotes the best model for each method.)}
\label{pallmonks}
\end{center}
\end{table}
% 
% \begin{figure}
% \nocaptionofalgo
% \caption{\textbf{test Update 3:} Gibbs sampling of the actor allocation vector $\K$ \label{alg:k}}
% \end{figure}

\begin{table}[htp]
\begin{center}
\begin{tabular}{| l|c|}
\hline
Update Type & Acceptance Rate (\%) \\
\hline
 Intercept ($\beta$) & 25.53 \\ 
 Latent Positions ($\Z$)&23.64\\ 
\hline
Allocation Updates ($\K$) & Acceptance Rate (\%) \\
\hline
 Gibbs update & -\\ 
 Move 1 & 1.89\\ 
 Move 2 & 15.61\\ 
 Move 3 & 1.62\\ 
Ejection& 3.99\\ 
Absorption & 3.99\\ 
 \hline
\end{tabular}
\caption{Acceptance rates for collapsed model sampling for Sampson's monks.}
\label{acceptmonks}
\end{center}
\end{table}

The results of our analysis for Sampson's monks network are qualitatively similar to the inference using \texttt{latentnet}.
  However inference using the collapsed model results in a dramatic reduction in CPU time. 
Qualitatively different results were seen for the variational approximation using \texttt{VBLPCM} 
with less separation of clusters and practically no uncertainty in cluster membership. 
Perhaps this is due to the approximation of the posterior distribution by the variational posterior. A major drawback of the variational approach is that the divergence 
between the two distributions can only be quantified up to an unknown constant of proportionality.

\subsection{Zachary's Karate Club}
% ~/Dropbox/phd triona/13febdraftpaper/Zachary12apr
Zachary's karate club  \cite{zachary77} consists of $78$ undirected friendship ties between $34$ members of  a karate club. 
The club split due to a disagreement between the club president and the coach, both of whom are included in the network as actors $1$ and $34$ respectively. 
The coach formed a new club with some of the members.
It is interesting to compare the actual split of the club and the clustering of the friendship network.
This is another example of a dynamic network which is usually examined in a static aggregated form in the social network analysis literature. 

\begin{figure}[htp]
 \begin{center}
\includegraphics[width=7cm]{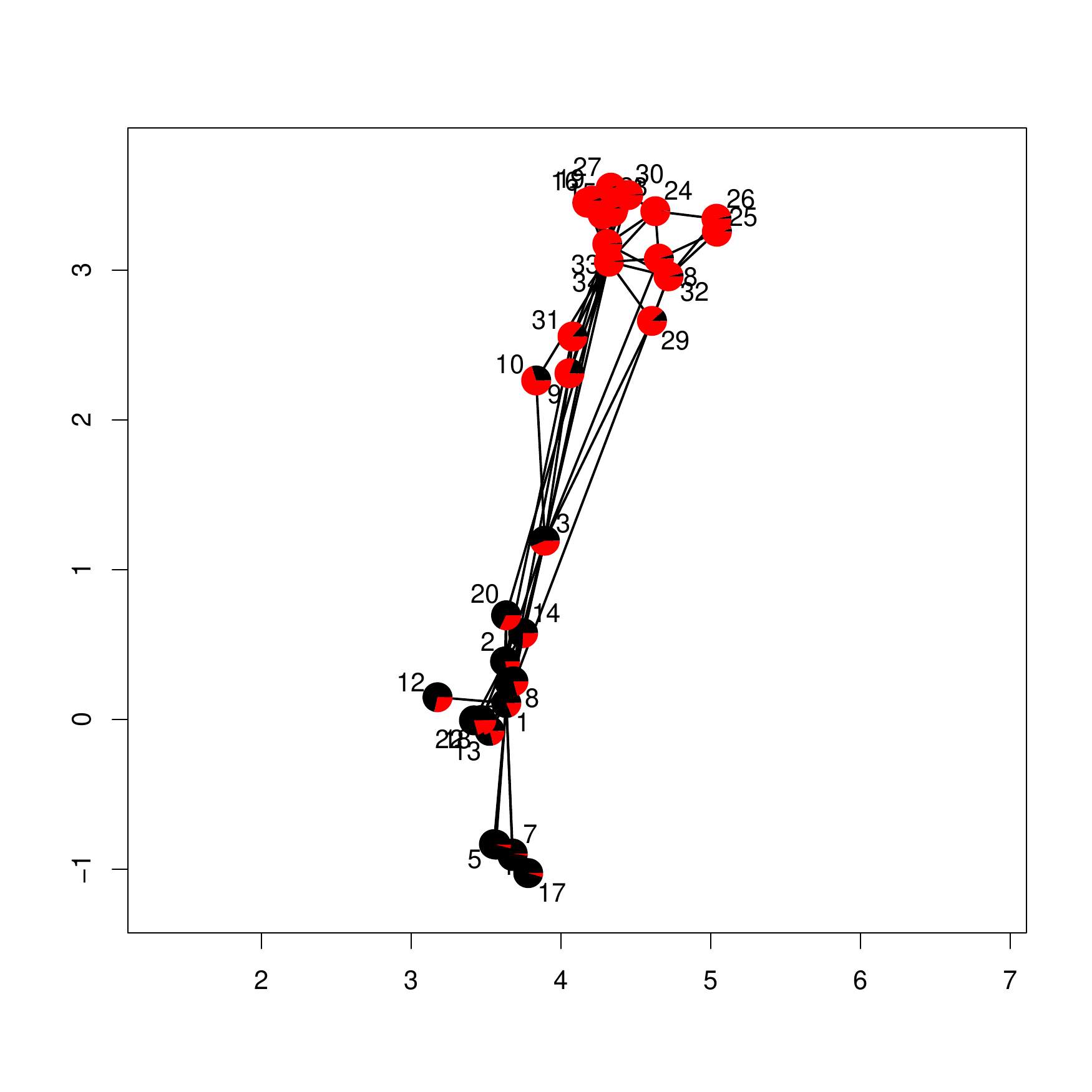}
\includegraphics[width=7cm]{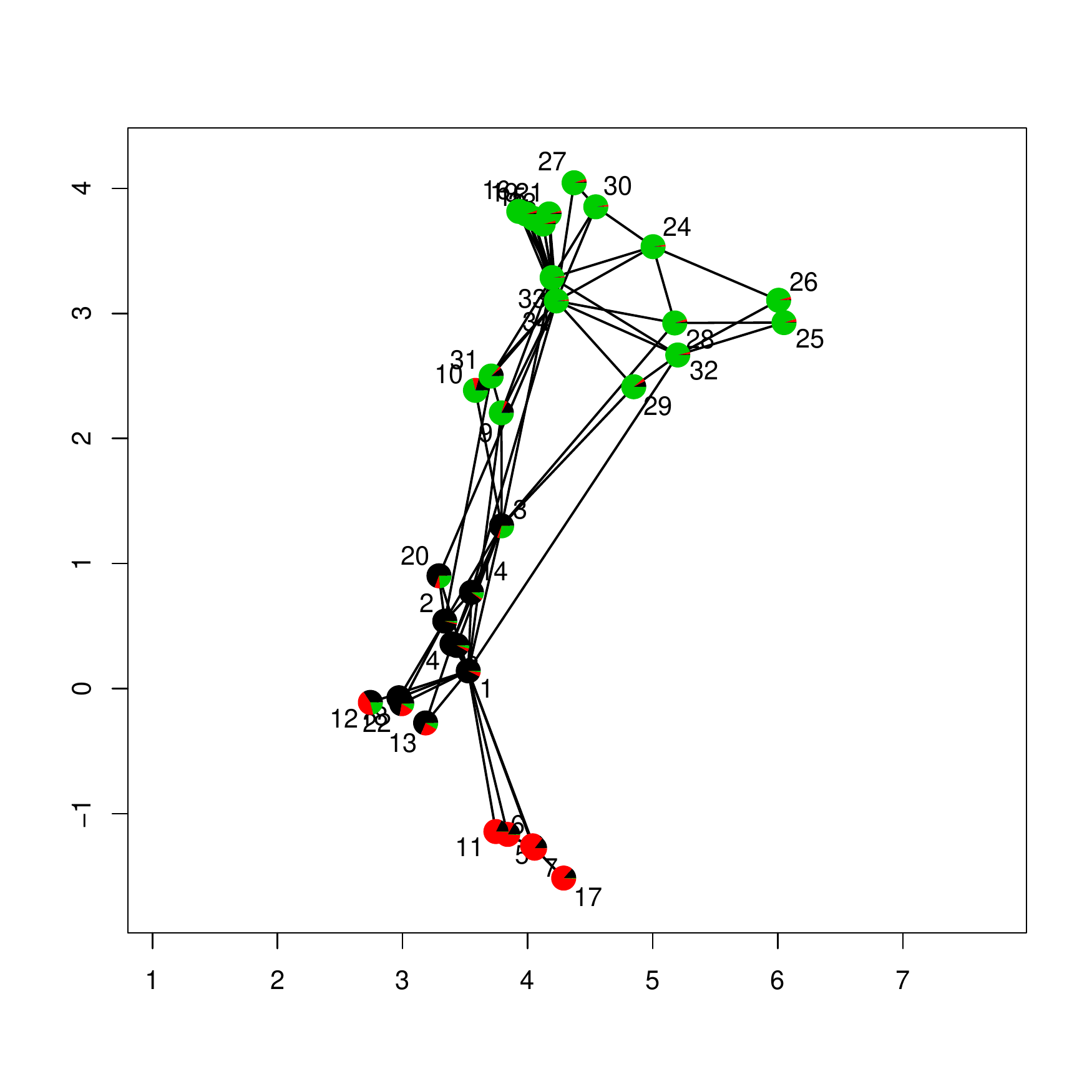}
 \end{center}
 \caption{Zachary's karate club posterior mean actor positions using the collapsed sampler for the most probable  $2$ and $3$ group models with a pie chart depicting uncertainty of cluster memberships.}% and latent positions using latent net for 2 group model}
\label{zmeanpiezac}
\end{figure}

\begin{figure}
        \centering
        \begin{subfigure}[b]{0.3\textwidth}
                \centering
                \includegraphics[width=\textwidth]{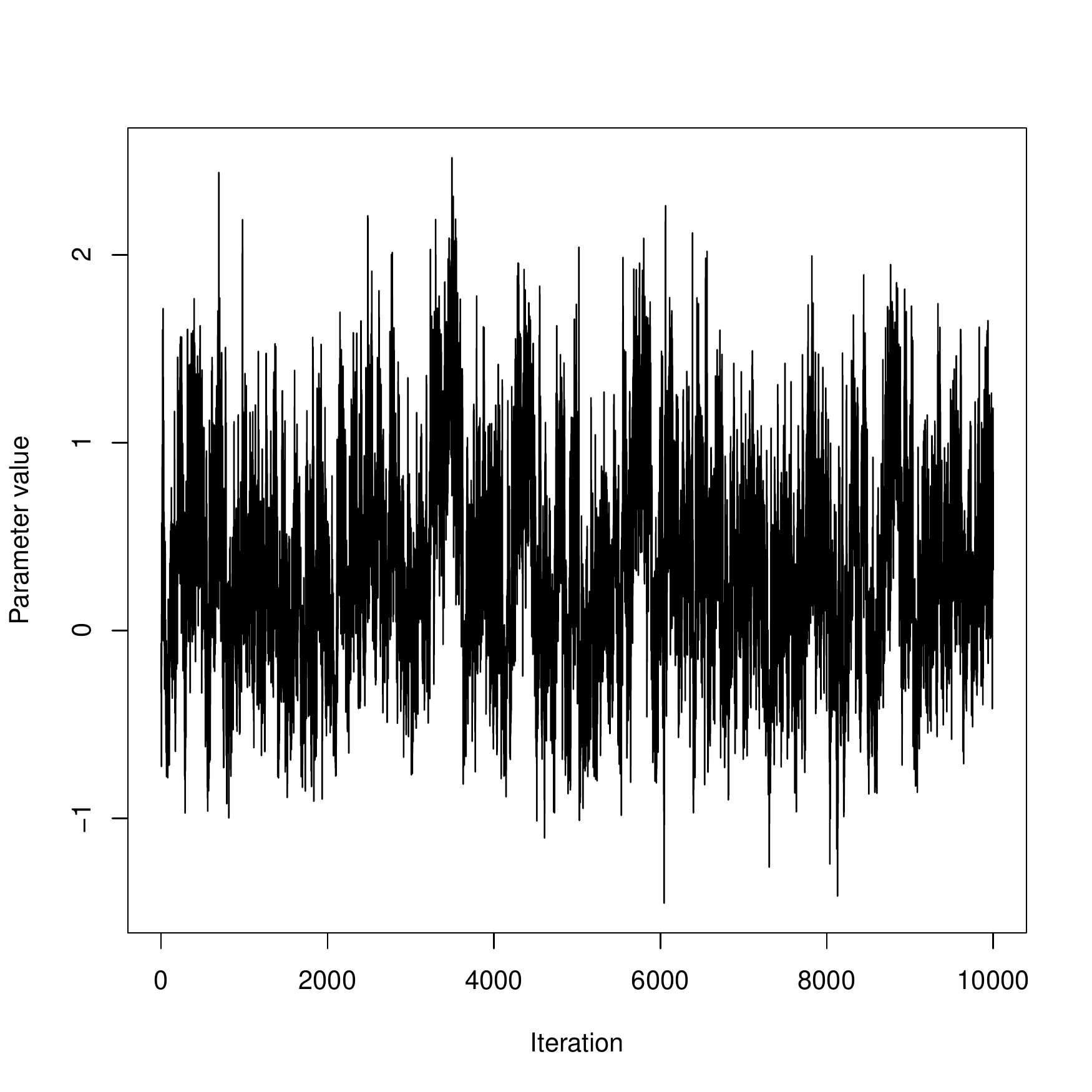}
                \caption{}\label{zbtr}
        \end{subfigure}%
%         \begin{subfigure}[b]{0.3\textwidth}
%                 \centering
%                 \includegraphics[width=\textwidth]{Zac15apr/gtrace.pdf}
%                 \caption{}\label{gtr}
%         \end{subfigure}%
          \begin{subfigure}[b]{0.3\textwidth}
                \centering
                \includegraphics[width=\textwidth]{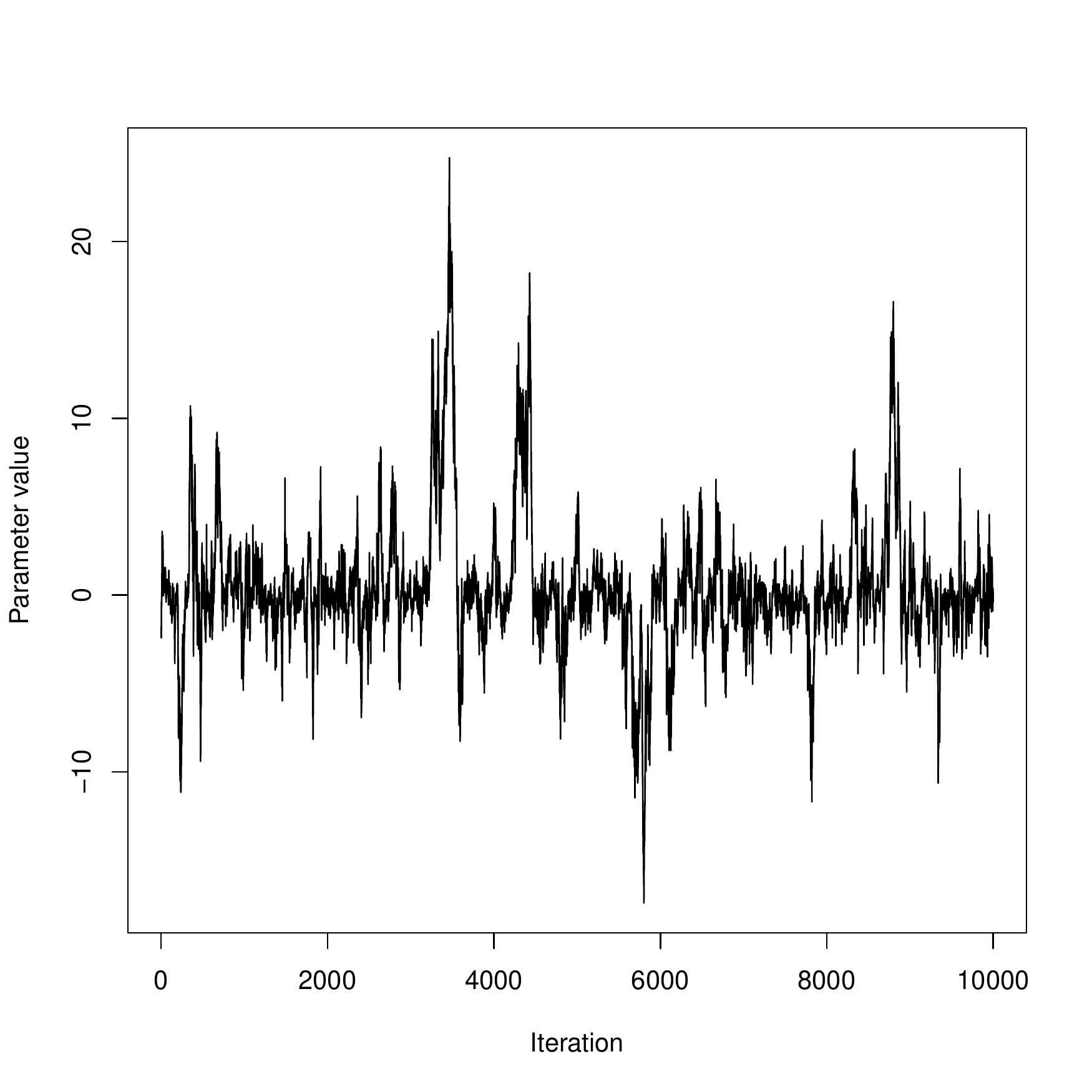}
                \caption{}
        \end{subfigure}
          \begin{subfigure}[b]{0.3\textwidth}
                \centering
                \includegraphics[width=\textwidth]{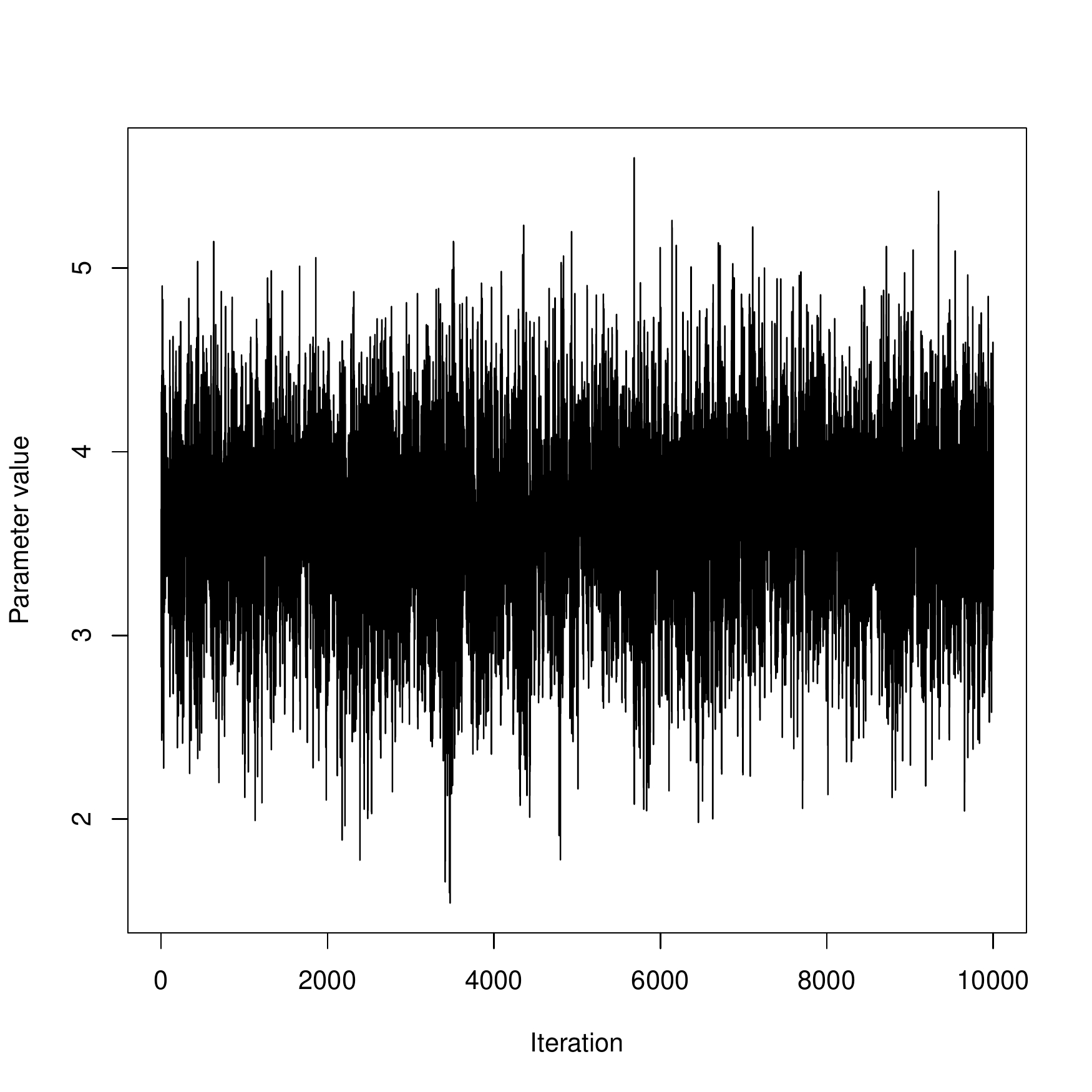}
                \caption{}
         \end{subfigure}
\caption{Trace plots of intercept $\beta$ (plot (a)) and one sample latent position $z_{(1,1)}$ pre- and post-Procrustes matching (plots (b), (c) respectively) 
for Zachary's karate club.}
\label{tracezac}
\end{figure}

Using the collapsed inference, the $3$ component model was favoured by a small margin with probability $0.38$. 
Posterior model probabilities for the collapsed method and the approximate BIC  inferred by \texttt{latentnet} and \texttt{VBLPCM} are displayed in Table \ref{pallzac}.  
CPU time was $17$ minutes for $1,000,000$ samples drawn from the collapsed posterior, thinning by $100$. 
The latent positions and the intercept mixed well (Figure \ref{tracezac}) 
using proposal variances $\sigma^2_z=1.7$ and $\sigma^2_\beta=0.5$. 
However, moves $1$ and $3$ on the allocation vector were slower to mix than for the smaller monks network. Perhaps
this is due to the increased size of the allocation vector. Thus the discrete set of possible allocation vectors to search across is very large. 
% Prior hyperparameters were set to $\alpha=2$, $\delta=0.103$ and $\nu=3$ per Handcock \textit{et al} \citeyear{handcocketal07}. We set $\omega^2 = 10$. 
The same hyperparameters were used as in Section \ref{resmonks}.
The resulting posterior mean actor positions for the collapsed method are shown in Figure \ref{zmeanpiezac}. 
There is good agreement between the actual club split and the clustering of our friendship network  for the $2$ group model.  The only discrepancy is actor $9$. He is clustered with the president in our analysis of this friendship network with probability $0.79$. 
However in reality he stayed in the coach Mr Hi's karate club due to the fact that he was only three weeks away from a test for his black belt (master status) when the split in the club
 occurred \cite{zachary77}. Otherwise there is good agreement. Combining two clusters of the $3$ group model, mirrors the true split as before, again with the
discrepancy of actor $9$. The $2$ group model looks quite linear here perhaps suggesting that a one-dimensional latent space may be appropriate. 

\begin{figure}
 \centering
\includegraphics[width=7cm]{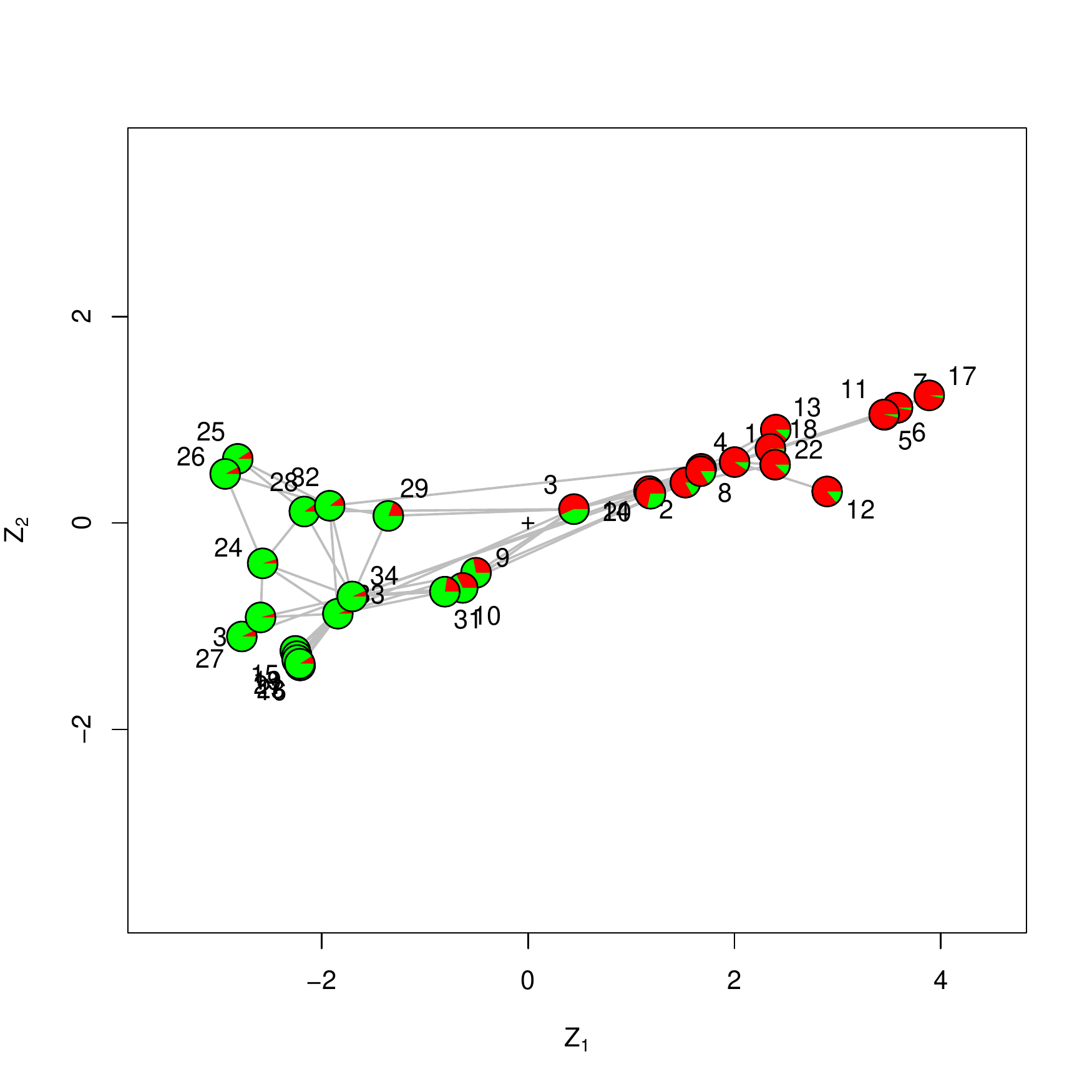}
\includegraphics[width=7cm]{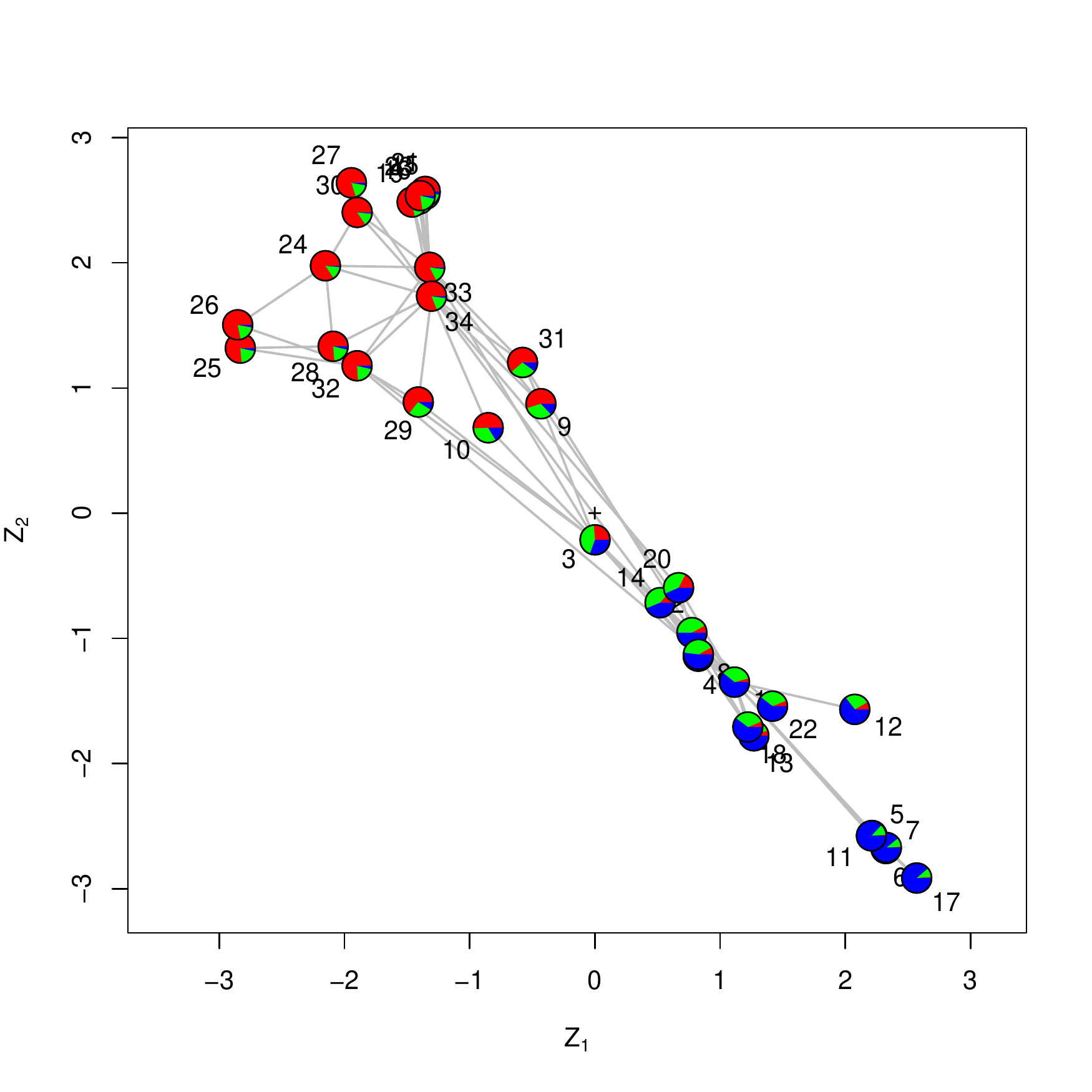}
\caption{Posterior mean latent positions for a $2$ and $3$ cluster models for Zachary's karate club using \texttt{latentnet}.}
\label{lnetzac}
\end{figure}

Inference using \texttt{latentnet} took $7$ to $13$ minutes to run  $100,000$ MCMC draws for each model. 
For up to maximum of $5$ groups, the full inference took approximately $50$ minutes. 
The \texttt{latentnet} position estimates are displayed in Figure \ref{lnetzac}.
There is good cluster agreement between the $2$ component model inferred using \texttt{latentnet} and the $2$ component model inferred using the collapsed sampler.
However, the $3$ component model inferred by the collapsed sampler displays less uncertainty in group membership than the corresponding estimate using  
\texttt{latentnet}. 
Actor positions are qualitatively similar for the $2$ component model but somewhat different for the $3$ component model.
% Allocations are different though the positions are qualitatively similar. 
% There is some concern about mixing here with high dependence factors in the Raftery and Lewis diagnostic checks. 
A possible influence is the difference in priors used in the methods.

\begin{figure}
 \centering
 \includegraphics[width=7cm]{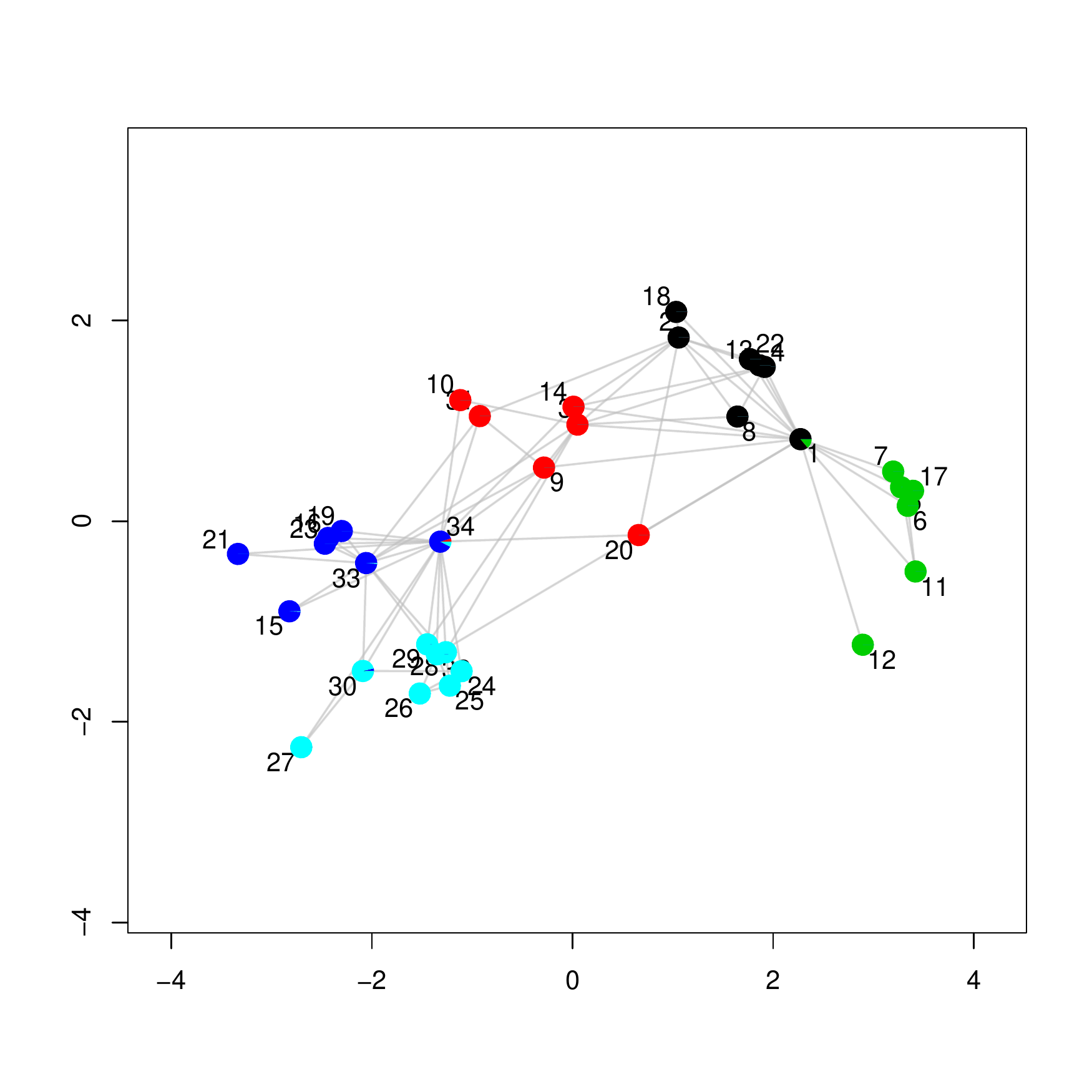} 
 \includegraphics[width=7cm]{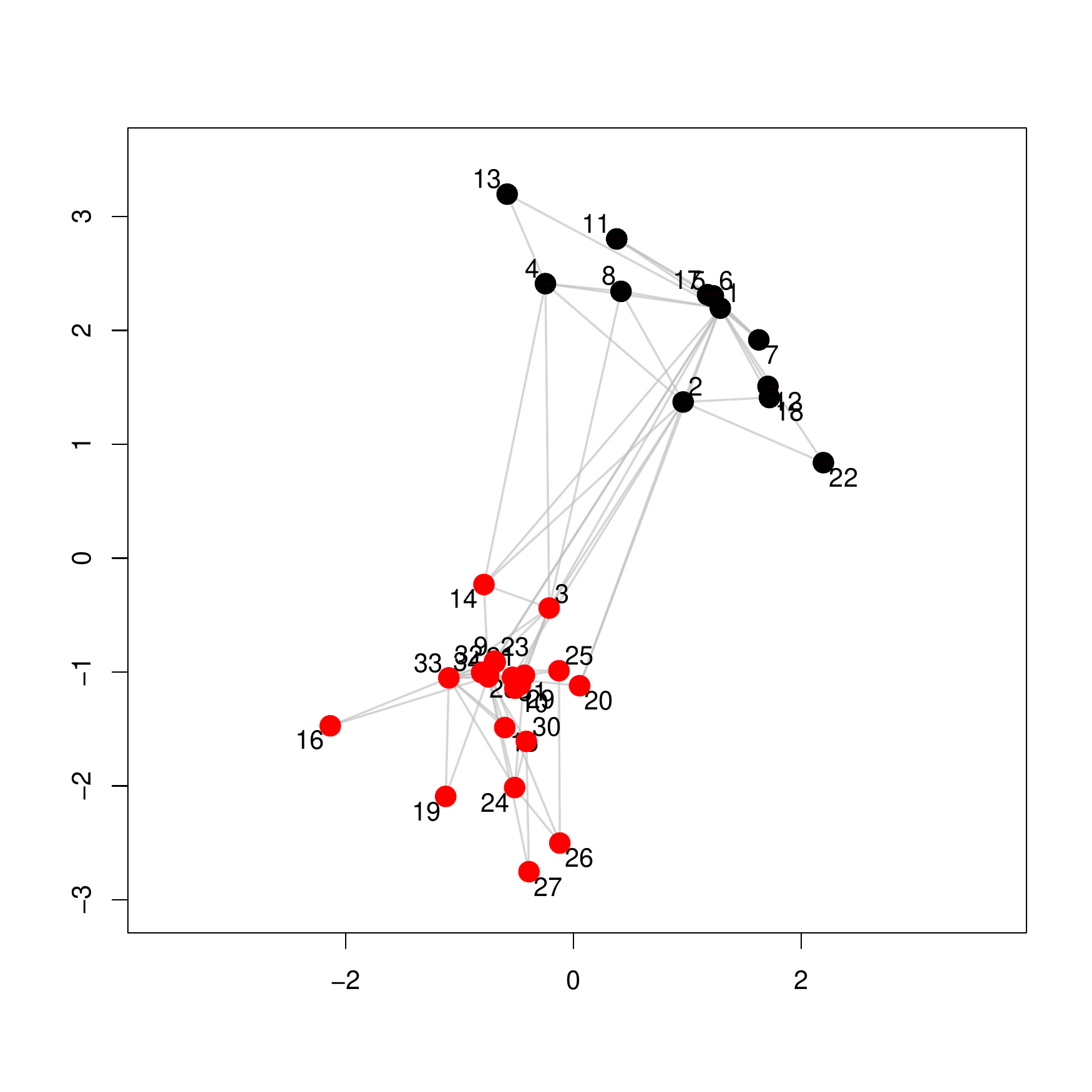} 
\caption{Variational Bayesian estimates of Latent positions of the $5$ and $2$ component model for Zachary's karate club.}
\label{vbzac}
\end{figure}

 %4.011132e-07 1.722132e-01 7.268594e-01 1.008600e-01 6.712042e-05
Using the variational Bayes package \texttt{VBLPCM} for $10$ runs of up to $5$ components took $7$ minutes and favoured the $5$ group model (Figure \ref{vbzac}). 
It is very much a hard clustering with almost no uncertainty in actor allocations which may suggest problems with using Variational Bayesian methods on this network.

\begin{table}[htp]
\begin{center}
\begin{tabular}{| l|c|c|c|c|c|}
\hline
 & $G=1$ & $G=2$ & $G=3$ & $G=4$ & $G=5$ \\
\hline
 Collapsed model probabilities $\pi(G|\Y)$& 0.2365 &   0.2807&   \underline{0.3769}&   0.0885&   0.0147\\ 
  \texttt{latentnet} BIC & 776.21 & \underline{747.40} & 750.63 &756.84 &770.64\\ 
%  \texttt{VBLPCM} BIC & 1241.21& 1126.49 &1140.06& 1126.56 &\underline{1124.60}\\
 \texttt{VBLPCM} BIC &1267.67 & 1134.30& 1109.42& 1093.94& \underline{1092.48}\\
 \hline
\end{tabular}
\caption{Posterior model probabilities for Zachary's karate club data using the collapsed algorithm and BIC values using \texttt{latentnet} and \texttt{VBLPCM}.
(The model underlined denotes the best model for each method.)}
\label{pallzac}
\end{center}
\end{table}

\begin{table}[htp]
\begin{center}
\begin{tabular}{| l|c|}
\hline
Update Type & Acceptance Rate (\%) \\
\hline
 Intercept ($\beta$) & 23.12 \\ 
 Latent Positions ($\Z$)&27.28\\ 
\hline
Allocation Updates ($\K$) & Acceptance Rate (\%) \\
\hline
 Gibbs update & -\\ 
 Move 1 & 0.66\\ 
 Move 2 & 7.50\\ 
 Move 3 & 0.54\\ 
Ejection& 1.41\\ 
Absorption & 2.23\\ 
 \hline
\end{tabular}
\caption{Acceptance rates (\%) for Zachary's karate club.}
\label{acceptzac}
\end{center}
\end{table}

The results are qualitatively similar for the collapsed method and the \texttt{latentnet} $2$ group model. 
Inference using the collapsed algorithm reduces computation time by at least
a factor of $2$, compared to inference using \texttt{latentnet}, even though the collapsed sampler has used many more iterations than \texttt{latentnet}. 
The \texttt{VBLPCM} algorithm was faster but chose a $5$ component model with practically no uncertainty in cluster membership. 

\subsection{Dolphin Network}

The dolphin network studied by Lusseau \textit{et al} \citeyear{lusseauetal03} represents social associations between $62$ dolphins living off Doubtful Sound in New Zealand. 
It is an undirected graph with $159$ ties.

\begin{figure}[htp]
 \begin{center}
\includegraphics[width=7cm]{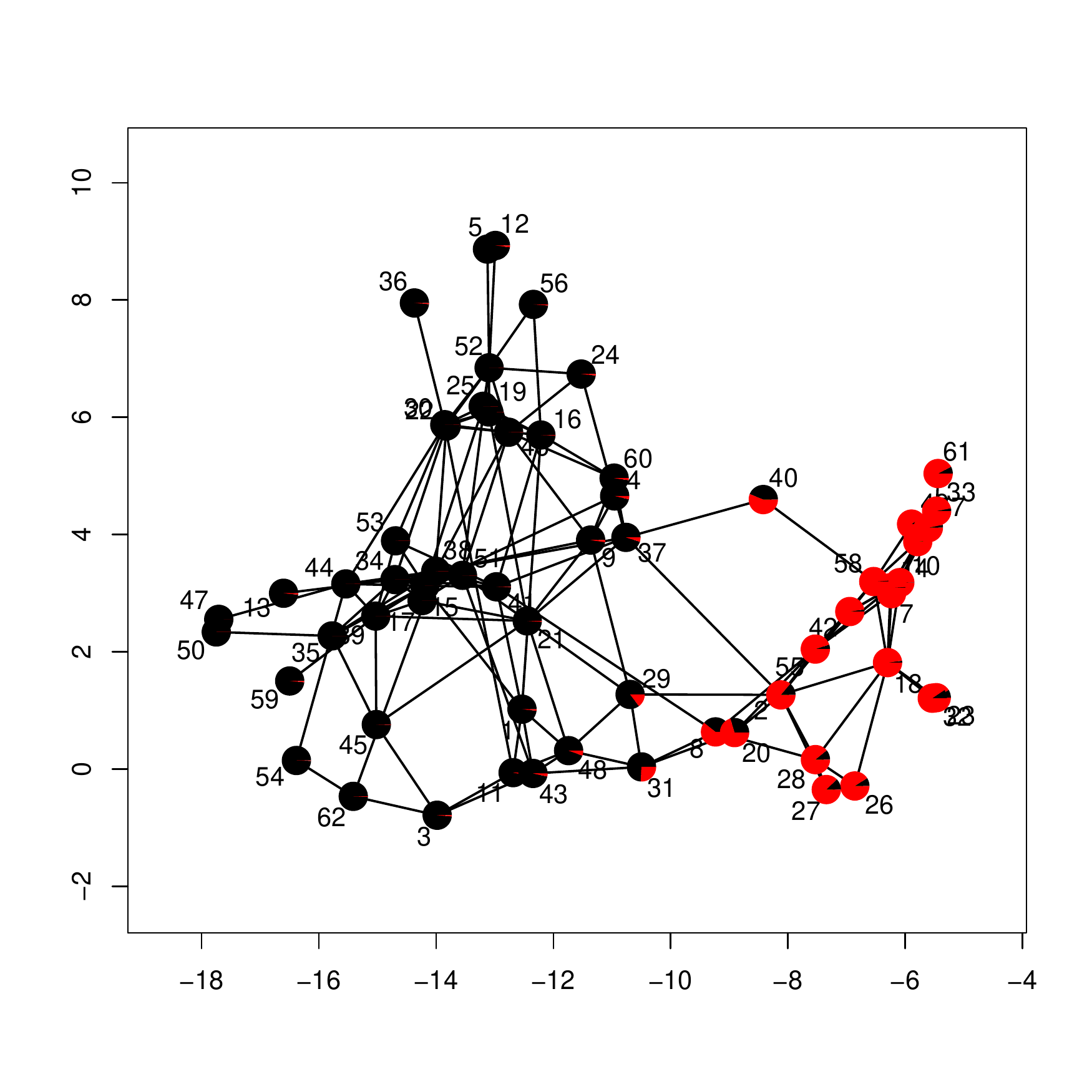}
 \end{center}
 \caption{The dolphin network's posterior mean actor positions using the collapsed sampler for the most probable $2$ group model with a pie chart depicting uncertainty of cluster memberships.}% and latent positions using latent net for 2 group model}
\label{zmeanpiedolph}
\end{figure}

\begin{figure}[htp]
 \centering
\includegraphics[width=7cm]{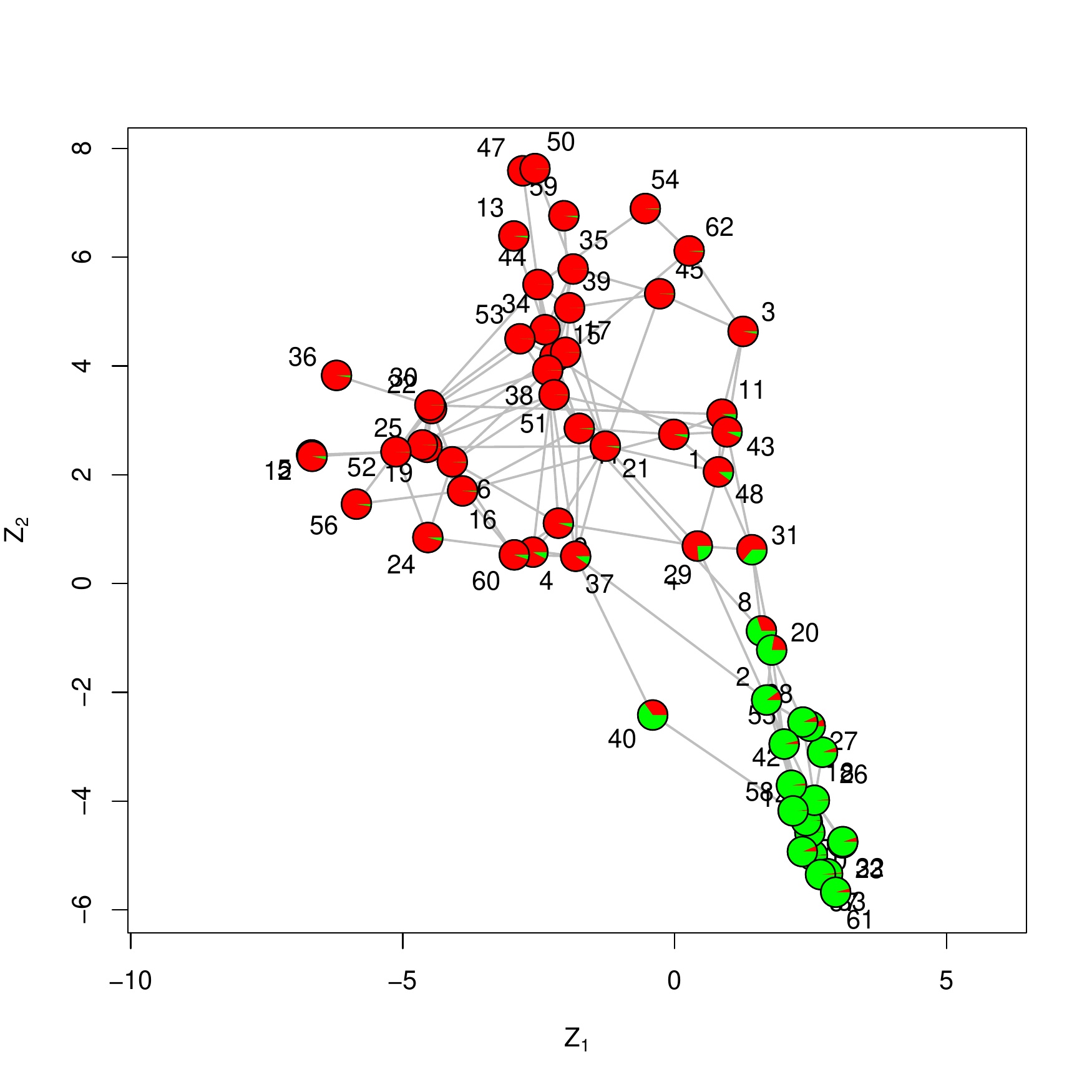}v
\caption{Posterior mean \texttt{latentnet} positions and uncertain clustering for the $2$ component model of the dolphin network.}
\label{lnetdolph}
\end{figure}

\begin{figure}
 \centering
\includegraphics[width=7cm]{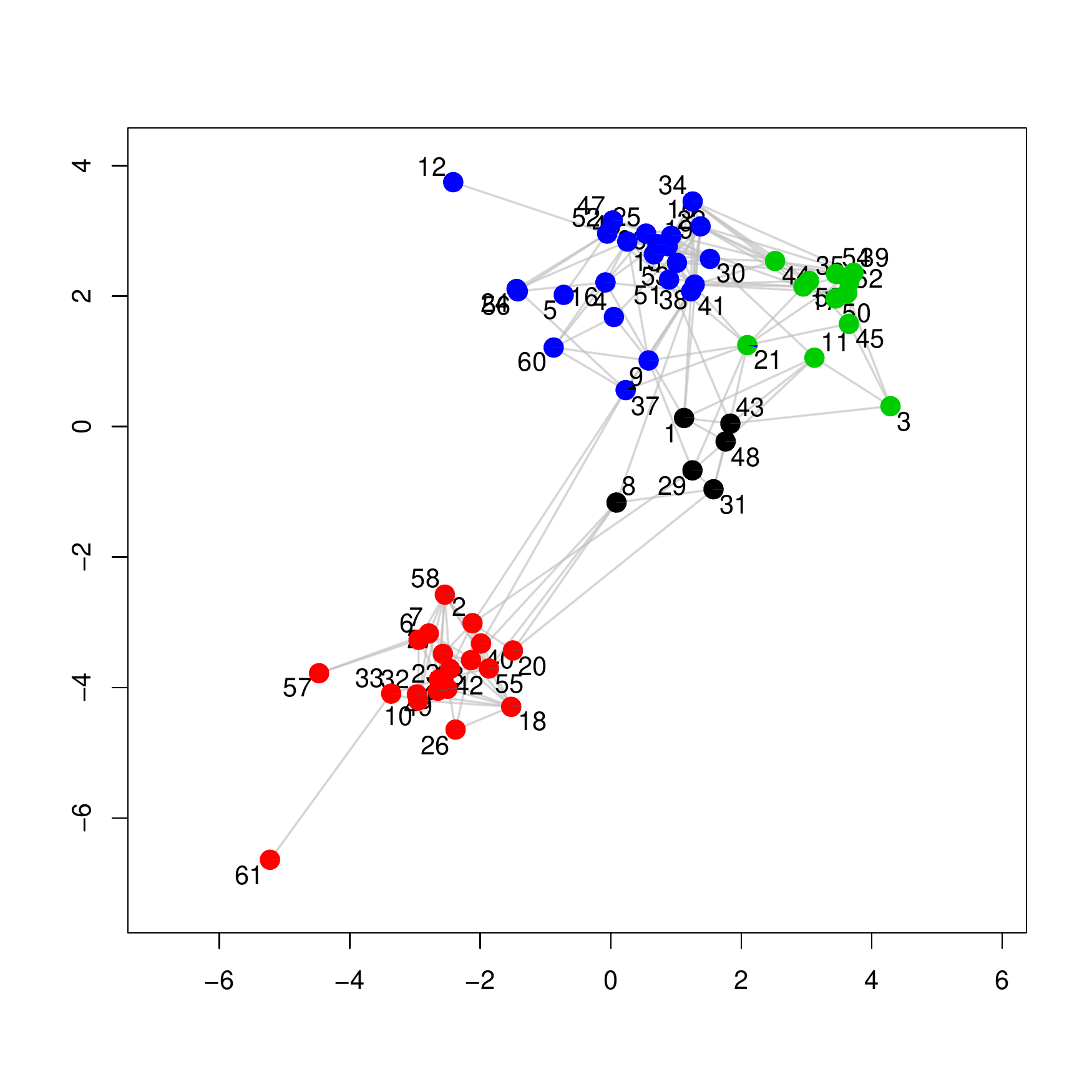} 
\caption{\texttt{VBLPCM} estimates of Latent positions for the $5$ component model for the dolphin network.}
\label{vbdolph}
\end{figure}

The $2$ group model was favoured by the collapsed allocation sampler and inference using \texttt{latentnet}. 
Posterior model probabilities for the collapsed method are displayed in Table \ref{palldolph} together
with the inferred BIC approximations to the approximated model evidence using \texttt{latentnet} and \texttt{VBLPCM}.  
In total $1,000,000$ draws of the collapsed posterior took 
$48$ minutes to run compared to $5$ hours for \texttt{latentnet} with $100,000$ MCMC draws of the posterior for
$5$ fitted models. Resulting positions and allocations are similar. 
The $5$ group model was chosen using \texttt{VBLPCM} which took $40$ minutes for $10$ variational fits of up to $5$ component models. 
The variational fit is displayed in Figure \ref{vbdolph} and is very different from both \texttt{latentnet} (Figure \ref{lnetdolph}) and our collapsed method
(Figure \ref{zmeanpiedolph}).

\begin{figure}
        \centering
        \begin{subfigure}[b]{0.3\textwidth}
                \centering
                \includegraphics[width=\textwidth]{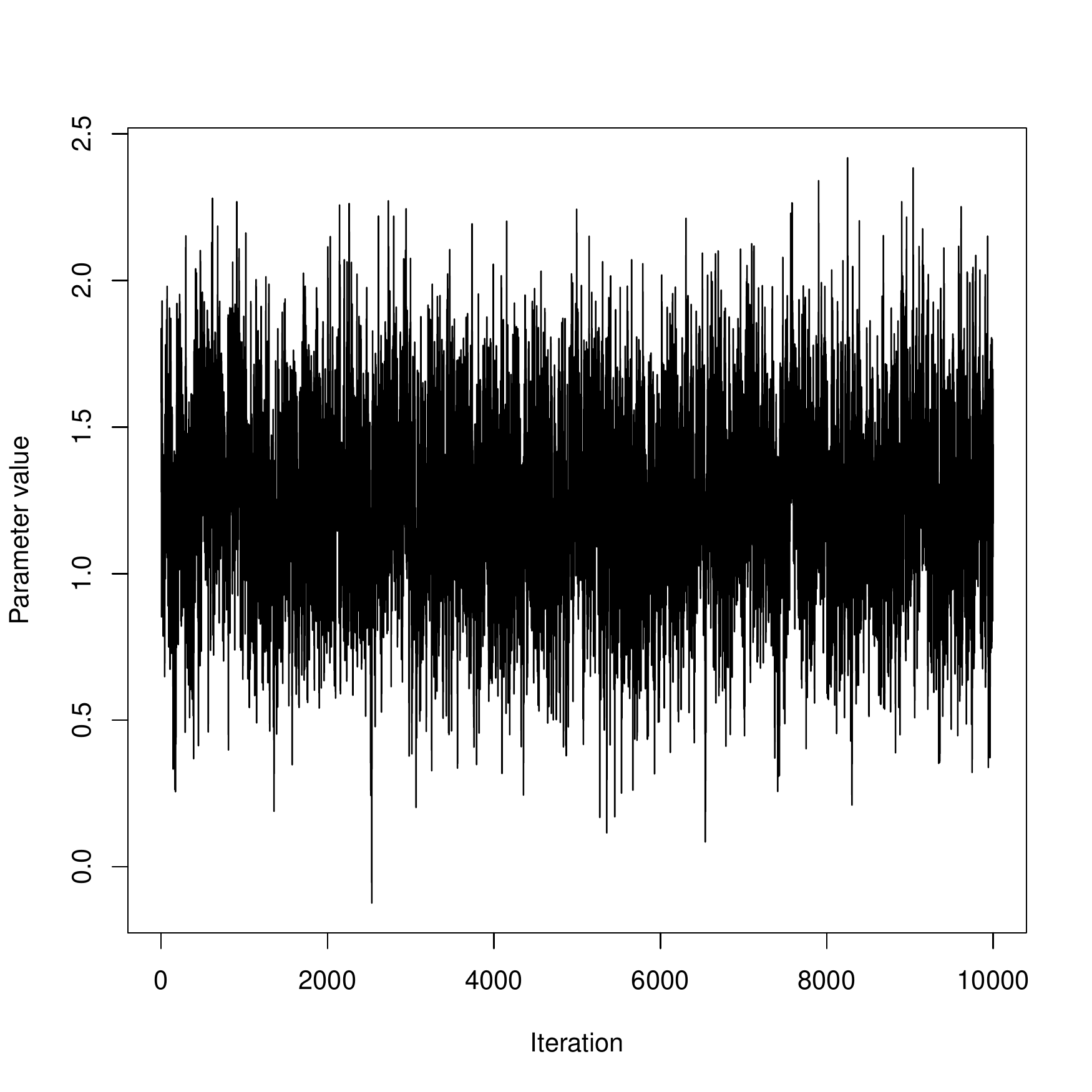}
                \caption{}\label{dbtr}
        \end{subfigure}%
%         \begin{subfigure}[b]{0.3\textwidth}
%                 \centering
%                 \includegraphics[width=\textwidth]{Zac15apr/gtrace.pdf}
%                 \caption{}\label{gtr}
%         \end{subfigure}%
              \begin{subfigure}[b]{0.3\textwidth}
                \centering
                \includegraphics[width=\textwidth]{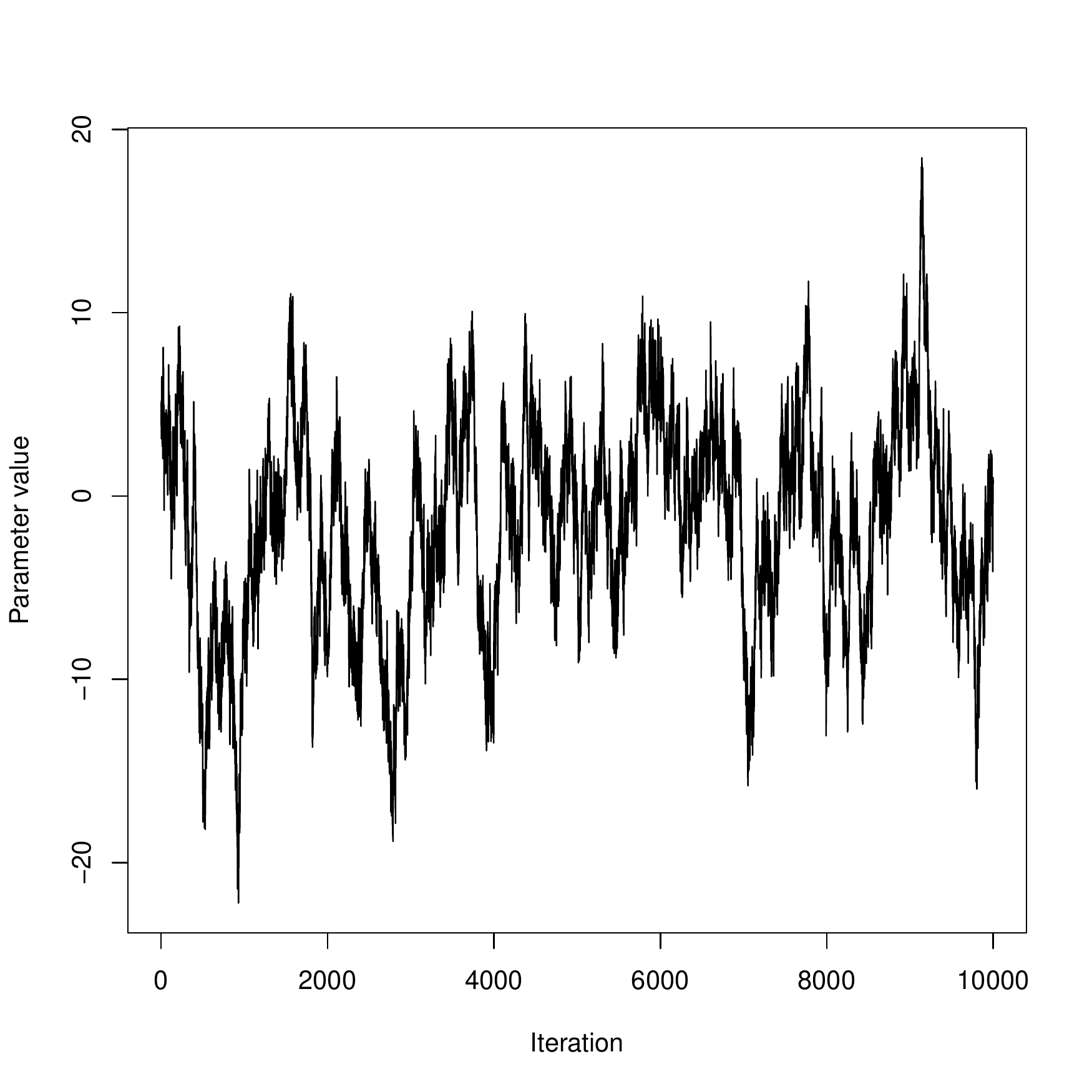}
                \caption{}
        \end{subfigure}
           \begin{subfigure}[b]{0.3\textwidth}
                \centering
                \includegraphics[width=\textwidth]{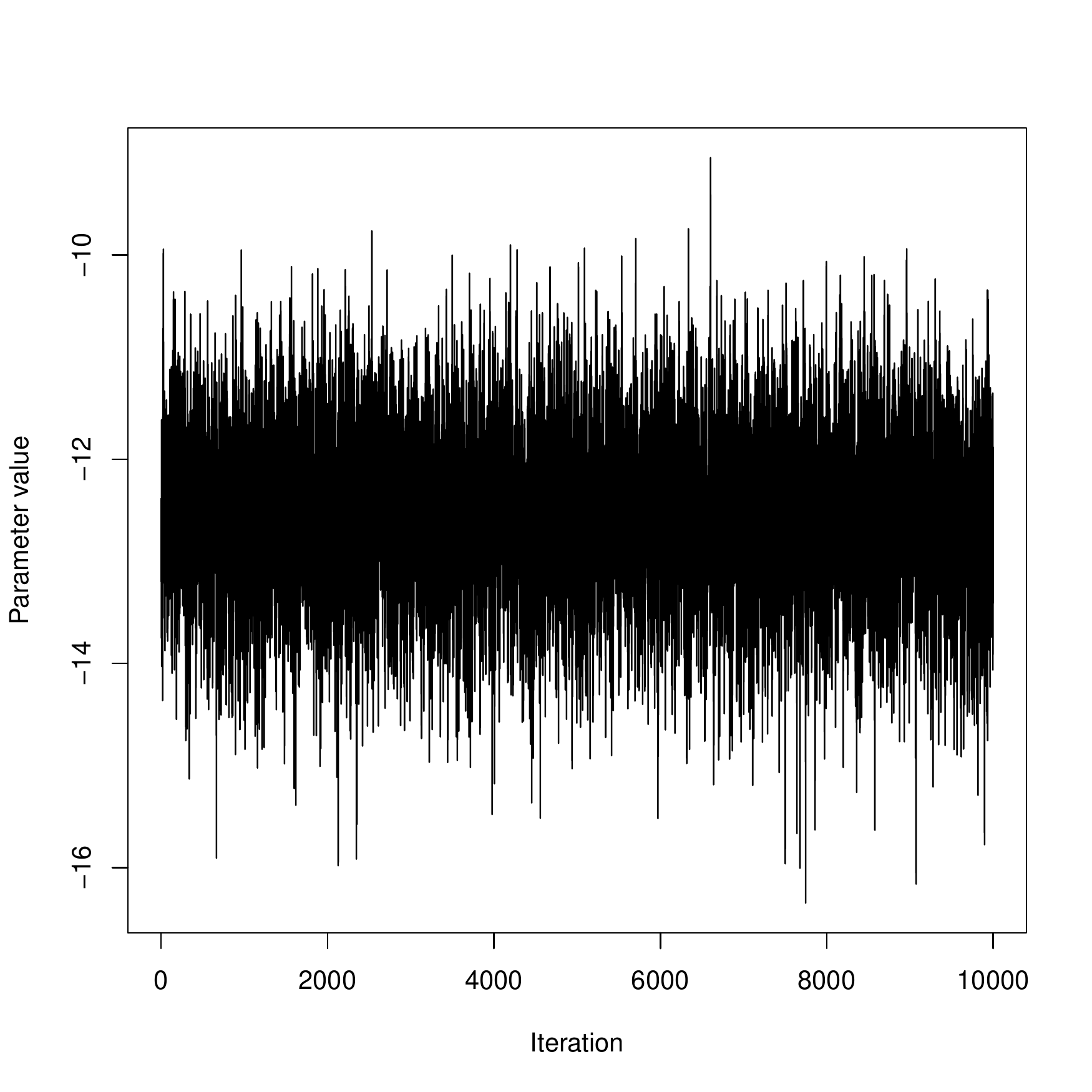}
                \caption{}
         \end{subfigure}
\caption{Trace plots of intercept $\beta$ (plot (a)) and one sample latent position $z_{(1,1)}$ pre- and post-Procrustes matching (plots (b), (c) respectively) 
for the dolphin network.}
\label{tracedolph}
\end{figure}

Posterior mean actor positions inferred by the collapsed sampling are displayed in Figure \ref{zmeanpiedolph}. 
As before, prior hyperparameters were set to $\alpha=2$, $\delta=0.103$, $\nu=3$ and $\omega^2 = 10$. 
Good mixing can be seen in Figure \ref{tracedolph} and acceptance rates are displayed in Table \ref{acceptdolph}. The chain was thinned by $100$.
  Proposal variances for the Metropolis-Hastings moves were $\sigma^2_z=3$ and $\sigma^2_{\beta}=0.2$ for the latent actor positions and for the intercept respectively. 

Good agreement can be seen between inference using \texttt{latentnet} and the collapsed sampler choosing the $2$ group model with qualitatively similar
 estimates of the latent actor positions as well as allocations. 
Results inferred by \texttt{VBLPCM} differed, favouring the $5$ group model with very little uncertainty in group membership.

\begin{table}[htp]
\begin{center}
\begin{tabular}{| l|c|c|c|c|c|}
\hline
 & $G=1$ & $G=2$ & $G=3$ & $G=4$ & $G=5$ \\
\hline
 Collapsed model probabilities $\pi(G|\Y)$&  0.0394 &   \underline{0.8986}&    0.0583&   0.0034&   0.0003\\ 
  \texttt{latentnet} BIC &1686.48&\underline{1660.16}&1667.23&1680.53&1690.32\\ 
%   \texttt{VBLPCM} BIC &2916.60 &2511.75 &2469.74& 2400.71& \underline{2387.89} \\
  \texttt{VBLPCM} BIC &3141.68 & 2537.09& 2506.26& 2449.28& \underline{2434.26}\\ 
  \hline
\end{tabular}
\caption{Posterior model probabilities for the dolphin network using the Collapsed algorithm and BIC using \texttt{latentnet} and \texttt{VBLPCM}
(The model underlined denotes the best model for each method.)}
\label{palldolph}
\end{center}
\end{table}
\begin{table}[htp]
\begin{center}
\begin{tabular}{| l|c|}
\hline
Update Type & Acceptance Rate (\%) \\
\hline
 Intercept ($\beta$) & 26.33 \\ 
 Latent Positions ($\Z$)&27.37\\ 
\hline
Allocation Updates ($\K$) & Acceptance Rate (\%) \\
\hline
 Gibbs update & -\\ 
 Move 1 & 0.04\\ 
 Move 2 & 2.73\\ 
 Move 3 & 0.09\\ 
Ejection& 0.34\\ 
Absorption & 0.38\\ 
 \hline
\end{tabular}
\caption{Acceptance rates (\%) for the dolphin network.}
\label{acceptdolph}
\end{center}
\end{table}

\section{Discussion} \label{discussion}
A novel approach to model selection for the latent position cluster model for social networks has been presented. 
Integrating out most of the clustering parameters from the model analytically provides a fixed dimensional parameter space for trans-model inference, 
allowing joint inference on the number of clusters in the network.
It avoids multiple approximations used by Handcock \textit{et al} \citeyear{handcocketal07} to estimate the model evidence,
while simultaneously improving computational efficiency compared with standard methods.
% Parallelisation is possible \ul{for the likliehood,} but not exploited in this paper and could contribute a further decrease in computational time.
Parallelisation is possible for the likelihood, but not exploited in this paper and could give further
decreases in computation time.
Mixing can be poor for the allocation vector using the collapsed sampler but despite this runs faster than \texttt{latentnet}.
 On the other hand, collapsed sampling is coded in $C$ whereas 
\texttt{latentnet} uses $R$ and $C$. Analysis for \texttt{latentnet} visits all models in turn whereas the collapsed sampling is done in one chain.

Our methodology was demonstrated using three real data examples with comparisons to current methods. 
Similar results were found between our methods and sampling the full posterior for separate models using \texttt{latentnet}.
Substantial uncertainty in the number of clusters and cluster membership was evident. 

\section{Appendix}
\subsection{Analytic integration of the clustering parameters from the latent position cluster model}
The collapsing of the latent position cluster model is detailed here. The clustering parameters \newline $\bftheta=(\bfmu,\bftau=1/\bfsigma^2,\bflambda)$ are integrated out of the model analytically.
The collapsed posterior distribution can be written as
\begin{eqnarray*}\label{collpost}
\pi(\Z,  \beta, \K, G|\Y) &=& \int_{\bftau}\int_{\bfmu}\int_{\bflambda} 
L(\Y|\Z,\beta) \pi(\Z|\bfmu, \bftau,\K,G)\pi(\K | \bflambda,G)\pi(\bflambda|G)
\pi(\bfmu|\bftau,G)\nonumber \\ 
&&\pi(\bftau|G) \pi(\beta) \pi(G)
\, \mbox{d}\bflambda\, \mbox{d}\bfmu \, \mbox{d}\bftau  \nonumber \\
&=& L(\Y|\Z,\beta) \pi(\beta) \pi(G)
\int_{\bftau}\pi(\bftau|G)\int_{\bfmu} \pi(\Z|\bfmu, \bftau,\K,G)\pi(\bfmu|\bftau,G) \nonumber \\ 
&&\int_{\bflambda} \pi(\K | \bflambda,G) \pi(\bflambda|G)
\, \mbox{d}\bflambda, \mbox{d}\bfmu \, \mbox{d}\bftau \ \nonumber \\
&=& L(\Y|\Z,\beta)  \pi(\beta) \pi(G)\int_{\bftau} \prod_{g=1}^G \frac{\left(\frac{\delta}{2}\right)^{\frac{\alpha}{2}} }{\Gamma\left(\frac{\alpha}{2}\right)}\tau_g^{\frac{\alpha}{2}-1}\exp\left\{-\frac{\delta}{2}\tau_g\right\}  \nonumber \\
&&  \int_{\bfmu}  \prod_{i=1}^N \prod_{g=1}^G\left( 
{\left( \frac{2\pi}{\tau_g} \right)}^{-d/2} \exp \left\{-\frac{\tau_g}{2}(\mathbf z_i- \bfmu_g)^T \I_d (\mathbf z_i-\bfmu_g) \right\}  
\right)^{\mathbbm{1}(k_i=g)} \nonumber\\
&&  \prod_{g=1}^G{\left( \frac{2\pi\omega^2}{\tau_g} \right)}^{-d/2} \exp \left\{-\frac{\tau_g}{2\omega^2}( \bfmu_g^T \I_d \bfmu_g) \right\} \,
\nonumber\\
&&
\int_{\bflambda}
\prod_{g=1}^G\lambda_g^{n_g}
\frac{\Gamma\left(G\nu\right)}{\Gamma\left(\nu\right)^G}\prod_{g=1}^G \lambda_g^{\nu-1}
\mbox{d}\bflambda,\mbox{d}\bfmu \, \mbox{d}\bftau \, ,  \nonumber\\
\end{eqnarray*}
where $n_g=\sum_{i=1}^n \mathbbm{1}(k_i=g)$ and we can rearrange to get, 
\begin{eqnarray*}
\pi(\Z,  \beta, \K, G|\Y)&=& L(\Y|\Z,\beta)  \pi(\beta) \pi(G)\frac{\Gamma\left(G\nu\right)}{\Gamma\left(\nu\right)^G} (2\pi)^{-\frac{d}{2}(n+G)} \frac{\left(\frac{\delta}{2}\right)^{\frac{G\alpha}{2}} }{\Gamma\left(\frac{\alpha}{2}\right)^G}
(\omega^2)^{-\frac{Gd}{2}} \nonumber\\
&& \int_{\bftau} \prod_{g=1}^G \tau_g^{\frac{1}{2}(n_gd+d+\alpha-2)}\exp\left\{-\frac{\delta}{2}\tau_g\right\} \nonumber\\
&& \int_{\bfmu} \prod_{g=1}^G  \exp \left\{-\frac{\tau_g}{2}\left(\sum_{i:k_i=g} \|\mathbf z_i- \bfmu_g\|^2 +\frac{1}{\omega^2}\|\bfmu_g\|^2\right)\right\}  
\nonumber\\ 
&&\int_{\bflambda}\prod_{g=1}^G \lambda_g^{n_g +\nu-1}
\, \mbox{d}\bflambda
\, \mbox{d}\bfmu 
\, \mbox{d}\bftau 
. \nonumber \\
\end{eqnarray*}
The mixing weights $\bflambda$ are collapsed or integrated out of this expression using the Dirichlet density where,
\begin{eqnarray*}
 \int_{\bflambda}\prod_{g=1}^G \lambda_g^{n_g+\nu-1}\, \mbox{d}{\bflambda}=\frac{\prod_{g=1}^G \Gamma\left(n_g+\nu\right)}{\Gamma\left(\sum_{g=1}^G(n_g+\nu)\right)}
=\frac{\prod_{g=1}^G \Gamma\left(n_g+\nu\right)}{\Gamma\left(n+G\nu\right)}.
\end{eqnarray*}
The cluster means $\bfmu$ are collapsed using Multivariate Normal densities,
with mean $\frac{\sum_{i:k_i=g} \mathbf z_i}{\left(n_g+\frac{1}{\omega^2}\right)}$ and covariance matrix
$\left(\tau_g\left(n_g+\frac{1}{\omega^2}\right)\I \right)^{-1}$,
\begin{eqnarray*}
\int_{\bfmu} \prod_{g=1}^G  \exp &&
\left\{-\frac{\tau_g}{2}\left(\sum_{i:k_i=g} \|\mathbf z_i- \bfmu_g\|^2 +\frac{1}{\omega^2}\|\bfmu_g\|^2\right)\right\} \nonumber\\ &&  
= \int_{\bfmu} \prod_{g=1}^G \exp \left\{-\frac{\tau_g}{2}\left(\sum_{i:k_i=g} (\mathbf \z_i^T\z_i- 2\mathbf z_i^T\bfmu_g+\bfmu_g^T\bfmu_g)+\frac{1}{\omega^2}\bfmu_g^T\bfmu_g\right)\right\}
\, \mbox{d}\bfmu \nonumber \\
&&=\int_{\bfmu} \prod_{g=1}^G \exp \left\{-\frac{\tau_g}{2}\left(\bfmu_g^T\bfmu_g\left(n_g+\frac{1}{\omega^2}\right)-2\sum_{i:k_i=g}  \mathbf z_i^T\bfmu_g +\sum_{i:k_i=g} \mathbf z_i^T \mathbf z_i\right)\right\}\, \mbox{d}{\bfmu}\\
&& =\int_{\bfmu}\prod_{g=1}^G\exp \left\{-\frac{\tau_g}{2}\left( \left(n_g+\frac{1}{\omega^2}\right)\left\|\bfmu_g-\frac{\sum_{i:k_i=g} \mathbf \z_i}{\left(n_g+\frac{1}{\omega^2}\right)}\right \|^2 
-\frac{\|\sum_{i:k_i=g} \mathbf z_i\|^2}{\left(n_g+\frac{1}{\omega^2}\right)} + \sum_{i:k_i=g} \|\mathbf z_i\|^2 \right)\right\} \, \mbox{d}\bfmu \\
&&=\prod_{g=1}^G { 2\pi }^{d/2} \left(\tau_g\left(n_g+\frac{1}{\omega^2}\right)\right)^{-\frac{d}{2}}
  \exp\left\{ -\frac{\tau_g}{2}  \left( \sum_{i:k_i=g} \|\mathbf z_i\|^2 - \frac{\|\sum_{i:k_i=g} \mathbf z_i\|^2}{\left(n_g+\frac{1}{\omega^2}\right)} \right)\right\}.
\end{eqnarray*}
Having integrated out $\bfmu$ and $\bflambda$ analytically, the resulting collapsed posterior is
\begin{eqnarray*}
\pi(\Z,  \beta, \K, G|\Y)
&=&L(\Y|\Z,\beta)  \pi(\beta) \pi(G)\frac{\Gamma\left(G\nu\right)}{\Gamma\left(\nu\right)^G} (2\pi)^{-\frac{d}{2}(n+G)} \frac{\left(\frac{\delta}{2}\right)^{\frac{G\alpha}{2}} }{\Gamma\left(\frac{\alpha}{2}\right)^G}
(\omega^2)^{-\frac{Gd}{2}}(2\pi)^{\frac{Gd}{2}}\prod_{g=1}^G \frac{\Gamma\left(n_g+\nu\right)}{\Gamma\left(n+G\nu\right)}\\
&\times& \int_{\bftau} \prod_{g=1}^G \left(n_g+\frac{1}{\omega^2}\right)^{-\frac{d}{2}}\tau_g^{\frac{1}{2}\left(dn_g+\alpha-2\right)}
\exp\left\{-\frac{\tau_g}{2}\left(\delta+\sum_{i:k_i=g} \|\mathbf z_i\|^2 - \frac{\|\sum_{i:k_i=g} \mathbf z_i\|^2}{\left(n_g+\frac{1}{\omega^2}\right)}\right) \right\}
  \, \mbox{d}\tau.\\
\end{eqnarray*}
The cluster precision parameter $\tau_g=1/\sigma_g^2$ is collapsed using Gamma($\alpha',\beta'$) densities, where $\alpha'=\frac{n_gd+\alpha}{2}$ and
 $\beta'=\frac{1}{2}\left(\sigma+\sum_{i:k_i=g} \|\mathbf z_i\|^2 - \frac{\|\sum_{i:k_i=g} \mathbf z_i\|^2}{\left(n_g+\frac{1}{\omega^2}\right)}\right)$, as follows,
\begin{eqnarray*}
\pi(\Z,  \beta, \K, G|\Y)
&=&L(\Y|\Z,\beta)  \pi(\beta) \pi(G)\frac{\Gamma\left(G\nu\right)}{\Gamma\left(\nu\right)^G} (2\pi)^{-\frac{d}{2}(n+G)} \frac{\left(\frac{\delta}{2}\right)^{\frac{G\alpha}{2}} }{\Gamma\left(\frac{\alpha}{2}\right)^G}
(\omega^2)^{-\frac{Gd}{2}}(2\pi)^{\frac{Gd}{2}}\prod_{g=1}^G \frac{\Gamma\left(n_g+\nu\right)}{\Gamma\left(n+G\nu\right)}\\
&\times& \prod_{g=1}^G \left(n_g+\frac{1}{\omega^2}\right)^{-\frac{d}{2}}\Gamma\left(\frac{n_gd+\alpha}{2}\right)
% \left(
  \left(\frac{1}{2}\left(
     \delta+\sum_{i:k_i=g} \|\mathbf z_i\|^2 - \frac{\|\sum_{i:k_i=g} \mathbf z_i\|^2}
  {\left(n_g+\frac{1}{\omega^2}\right)}\right)\right)
  ^{-\left(\frac{n_gd+\alpha}{2}\right)}.
\end{eqnarray*}
Finally, the fully collapsed posterior for the latent position cluster model, including expressions for the likelihood, the prior for $\beta$ and the prior on for the number of groups $G$ is
\begin{eqnarray*}
\pi(\Z,  \beta, \K, G|\Y)&=&
 \displaystyle\prod_{i=1}^n{} \displaystyle\prod_{j \neq i} \frac{\exp\{y_{ij} (\beta-||\mathbf z_i - \mathbf z_j||)\}}{1+\exp(\beta-||\mathbf z_i - \mathbf z_j||)}\\
&&\times \frac{1}{\sqrt {2\pi\psi  } }\exp\left\{-\frac{(\beta-\xi)^2}{2\psi} \right\} 
\frac{\exp\{-1\}}{G!}\\
&&\times \frac{\Gamma\left(G\nu\right)}{\Gamma\left(\nu\right)^G} \pi^{-\frac{dn}{2}}\frac{\left(\delta\right)^{\frac{G\alpha}{2}} }{\Gamma\left(\frac{\alpha}{2}\right)^G}
(\omega^2)^{-\frac{Gd}{2}}\frac{\prod_{g=1}^G \Gamma\left(n_g+\nu\right)}{\Gamma\left(n+G\nu\right)}\\
&&\times  \prod_{g=1}^G\frac{\Gamma\left(\frac{n_gd+\alpha}{2}\right)}{ \left(n_g+\frac{1}{\omega^2}\right)^{\frac{d}{2}}}\left(\delta+\sum_{i:k_i=g} \|\mathbf z_i\|^2 - \frac{\|\sum_{i:k_i=g} \mathbf z_i\|^2}{\left(n_g+\frac{1}{\omega^2}\right)}\right)^{-\left(\frac{n_gd+\alpha}{2}\right)}.
\end{eqnarray*}

% \subsection{Graphical model to illustrate parameter dependencies within the latent position cluster model}
% 
% LPCM
% \def \objectstyle{\hbox}
% \entrymodifiers={++[F-:<3pt>]}
% \centerline{
% \xymatrix@R=15pt{
% $\alpha$\ar[d] & *{} &$G$\ar[ddl]\ar[dll]\ar[ddr]\ar[ddd]&*{} \\
% $\bflambda$\ar[dr]&*{} &*{}&$\phi$\ar[d] \\%& $\phi$\ar[dl]&*{}\\
% *{} & $\K$\ar[dr] &*{} & $\bfmu, \sigma$ \ar[dl]\\
% *{}& $\beta$ \ar[dr] & $\Z$\ar[d] &*{} &*{} \\
%  *{} & *{} &$\Y$ &*{}
% }
% }
% 
% Collapsed LPCM
% \def \objectstyle{\hbox}
% \entrymodifiers={++[F-:<3pt>]}
% \centerline{
% \xymatrix@R=15pt{
% $\alpha$\ar[d] & *{} &$G$\ar[ddl]\ar[dll]\ar[ddr]\ar[ddd]&*{} \\
% REMOVE $\bflambda$\ar[dr]&*{} &*{}&$\phi$\ar[d] \\%& $\phi$\ar[dl]&*{}\\
% *{} & $\mathbf k$\ar[dr] &*{} & REMOVE $\bftheta$ \ar[dl]\\
% *{}& $\beta$ \ar[dr] & $\Z$\ar[d] &*{} &*{} \\
%  *{} & *{} &$\Y$ &*{}
% }
% }#

% \section{Notes for reference myself}
% 
% \begin{itemize}
%  \item Determinant $\tau I$ is $\tau^d$
%  \item $\|\X\|^2=X^TX$
% %  \item 
% \end{itemize}

\paragraph*{Acknowledgements:} Nial Friel and Caitr\'{i}ona Ryan's research was supported by a Science Foundation Ireland
Research Frontiers Program grant, 09/RFP/MTH2199. This research was also supported in part by a research grant from Science 
Foundation Ireland (SFI) under Grant Number SFI/12/RC/2289. Jason Wyse'`s research was supported through the STATICA project, a 
Principal Investigator program of Science Foundation Ireland, 08/IN.1/I1879.

\section*{Supplemental Materials}
\begin{description}
 \item[C code:] The supplemental files for this article include the C implementation of the algorithms of Section \ref{algs}. The examples
of Section \ref{res} can be reproduced by calling an R script also provided. 
Please see the file \texttt{README} contained within the accompanying tar file for more details.
%  \item[C code:] The supplemental files for this article include C programs which can be used to
% replicate the Ising model study and exponential random graph example in Section~\ref{sec:examples} of this article. 
% Please see the file \texttt{README.txt} contained within the accompanying tar file for more details. 
\end{description}

\bibliographystyle{mybib}
\bibliography{biblio20mayba}

\begin{thebibliography}{}

\bibitem[\protect\citeauthoryear{Adamic, Lukose, Puniyani and Huberman}{Adamic
  {\em et~al\/}}{2001}]{adamicetal01}
Adamic, L., R.~Lukose, A.~Puniyani and B.~Huberman (2001), Search in power-law
  networks. {\em Physical Review E\/} {\bf 64}(4), 046135

\bibitem[\protect\citeauthoryear{Breslow}{Breslow}{1996}]{breslow1996}
Breslow, N. (1996), Statistics in epidemiology: the case-control study. {\em
  Journal of the American Statistical Association\/} {\bf 91}(433), 14--28

\bibitem[\protect\citeauthoryear{Carpaneto and Toth}{Carpaneto and
  Toth}{1980}]{carpaneto1980algorithm}
Carpaneto, G. and P.~Toth (1980), Algorithm 548: Solution of the assignment
  problem [H]. {\em ACM Transactions on Mathematical Software (TOMS)\/} {\bf
  6}(1), 104--111

\bibitem[\protect\citeauthoryear{Dellaportas and Papageorgiou}{Dellaportas and
  Papageorgiou}{2006}]{dell:pap06}
Dellaportas, P. and I.~Papageorgiou (2006), Multivariate mixtures of normals
  with unknown number of components. {\em Statistics and Computing\/} {\bf
  16}(1), 57--68

\bibitem[\protect\citeauthoryear{Faloutsos, Faloutsos and Faloutsos}{Faloutsos
  {\em et~al\/}}{1999}]{faloutsosetal1999}
Faloutsos, M., P.~Faloutsos and C.~Faloutsos (1999), On power-law relationships
  of the internet topology. In {\em ACM SIGCOMM Computer Communication
  Review\/}, vol.~29, pp. 251--262, ACM

\bibitem[\protect\citeauthoryear{Fraley and Raftery}{Fraley and
  Raftery}{2002}]{fraley:raftery02}
Fraley, C. and A.~E. Raftery (2002), Model-based clustering, discriminant
  analysis, and density estimation. {\em Journal of the American Statistical
  Association\/} {\bf 97}(458), 611--631

\bibitem[\protect\citeauthoryear{Fraley and Raftery}{Fraley and
  Raftery}{2003}]{fraley:raftery03}
Fraley, C. and A.~E. Raftery (2003), Enhanced model-based clustering, density
  estimation, and discriminant analysis software: MCLUST. {\em Journal of
  Classification\/} {\bf 20}(2), 263--286

\bibitem[\protect\citeauthoryear{Friel and Wyse}{Friel and
  Wyse}{2012}]{friel:wyse12}
Friel, N. and J.~Wyse (2012), Estimating the evidence--a review. {\em
  Statistica Neerlandica\/} {\bf 66}(3), 288--308

\bibitem[\protect\citeauthoryear{Handcock, Raftery and Tantrum}{Handcock {\em
  et~al\/}}{2007}]{handcocketal07}
Handcock, M., A.~Raftery and J.~Tantrum (2007), Model-based clustering for
  social networks. {\em Journal of the Royal Statistical Society: Series A
  (Statistics in Society)\/} {\bf 170}(2), 301--354

\bibitem[\protect\citeauthoryear{Hoff and Handcock}{Hoff and
  Handcock}{2002}]{hoff:raft:hand02}
Hoff, P.D.and~Raftery, A. and M.~Handcock (2002), Latent space approaches to
  social network analysis. {\em Journal of the American Statistical
  Association\/} {\bf 97}(460), 1090--1098

\bibitem[\protect\citeauthoryear{Kolaczyk}{Kolaczyk}{2009}]{kolaczyk09}
Kolaczyk, E. (2009), {\em Statistical analysis of network data: methods and
  models\/}. Springer

\bibitem[\protect\citeauthoryear{Krivitsky and Handcock}{Krivitsky and
  Handcock}{2008}]{Kriv:Hand07}
Krivitsky, P. and M.~Handcock (2008), Fitting Latent Cluster Models for
  Networks with latentnet. {\em Journal of Statistical Software\/} {\bf 24}(5),
  1--23

\bibitem[\protect\citeauthoryear{Krivitsky and Handcock}{Krivitsky and
  Handcock}{2013}]{Kriv:Hand13}
Krivitsky, P.~N. and M.~S. Handcock (2013), {\em latentnet: Latent position and
  cluster models for statistical networks\/}. The Statnet Project
  (\url{http://www.statnet.org}), R package version 2.4-4

\bibitem[\protect\citeauthoryear{Lusseau, Schneider, Boisseau, Haase, Slooten
  and Dawson}{Lusseau {\em et~al\/}}{2003}]{lusseauetal03}
Lusseau, D., K.~Schneider, O.~Boisseau, P.~Haase, E.~Slooten and S.~Dawson
  (2003), The bottlenose dolphin community of Doubtful Sound features a large
  proportion of long-lasting associations. {\em Behavioral Ecology and
  Sociobiology\/} {\bf 54}(4), 396--405

\bibitem[\protect\citeauthoryear{Michailidis}{Michailidis}{2012}]{michailidis1%
2}
Michailidis, G. (2012), Statistical Challenges in Biological Networks. {\em
  Journal of Computational and Graphical Statistics\/} {\bf 21}(4), 840--855

\bibitem[\protect\citeauthoryear{Nobile}{Nobile}{2007}]{nobile07}
Nobile, A. (2007), Bayesian finite mixtures: a note on prior specification and
  posterior computation. {\em arXiv preprint arXiv:0711.0458\/}

\bibitem[\protect\citeauthoryear{Nobile and Fearnside}{Nobile and
  Fearnside}{2007}]{nob:fearn07}
Nobile, A. and A.~Fearnside (2007), Bayesian finite mixtures with an unknown
  number of components: the allocation sampler. {\em Statistics and
  Computing\/} {\bf 17}(2), 147--162

\bibitem[\protect\citeauthoryear{Nowicki and Snijders}{Nowicki and
  Snijders}{2001}]{nowicki:snijders01}
Nowicki, K. and T.~Snijders (2001), Estimation and prediction for stochastic
  blockstructures. {\em Journal of the American Statistical Association\/} {\bf
  96}(455), 1077--1087

\bibitem[\protect\citeauthoryear{Phillips and Smith}{Phillips and
  Smith}{1996}]{phil:smith96}
Phillips, D. and A.~Smith (1996), Bayesian model comparison via jump
  diffusions. {\em Markov chain Monte Carlo in practice\/} pp. 215--239

\bibitem[\protect\citeauthoryear{Raftery, Niu, Hoff and Yeung}{Raftery {\em
  et~al\/}}{2012}]{rafteryetal2012}
Raftery, A., X.~Niu, P.~Hoff and K.~Yeung (2012), Fast inference for the latent
  space network model using a case-control approximate likelihood. {\em Journal
  of Computational and Graphical Statistics\/} {\bf 21}(4), 901--919

\bibitem[\protect\citeauthoryear{Richardson and Green}{Richardson and
  Green}{1997}]{richardson:green97}
Richardson, S. and P.~Green (1997), On Bayesian Analysis of Mixtures with an
  Unknown Number of Components (with discussion). {\em Journal of the Royal
  Statistical Society: Series B (Statistical Methodology)\/} {\bf 59}, 731--792

\bibitem[\protect\citeauthoryear{Robins, Snijders, Wang, Handcock and
  Pattison}{Robins {\em et~al\/}}{2007}]{robinsetal07b}
Robins, G., T.~Snijders, P.~Wang, M.~Handcock and P.~Pattison (2007), Recent
  developments in exponential random graph (p*) models for social networks.
  {\em Social Networks\/} {\bf 29}(2), 192--215

\bibitem[\protect\citeauthoryear{Salter-Townshend and Murphy}{Salter-Townshend
  and Murphy}{2012}]{salter:murphy12}
Salter-Townshend, M. and T.~Murphy (2012), Variational Bayesian Inference for
  the Latent Position Cluster Model for network data. {\em Computational
  Statistics \& Data Analysis\/} {\bf 57}(1), 661--671

\bibitem[\protect\citeauthoryear{Sampson}{Sampson}{1968}]{sampson68}
Sampson, S. (1968), {\em A novitiate in a period of change: An experimental and
  case study of social relationships\/}. Ph.D. thesis, Cornell University,
  September

\bibitem[\protect\citeauthoryear{Schwarz}{Schwarz}{1978}]{schwarz1978}
Schwarz, G. (1978), Estimating the dimension of a model. {\em The Annals of
  Statistics\/} {\bf 6}(2), 461--464

\bibitem[\protect\citeauthoryear{Shortreed, Handcock and Hoff}{Shortreed {\em
  et~al\/}}{2006}]{shortreedetal06}
Shortreed, S., M.~Handcock and P.~Hoff (2006), Positional estimation within a
  latent space model for networks. {\em Methodology: European Journal of
  Research Methods for the Behavioral and Social Sciences\/} {\bf 2}(1), 24--33

\bibitem[\protect\citeauthoryear{Sibson}{Sibson}{1978}]{sibson1978}
Sibson, R. (1978), Studies in the robustness of multidimensional scaling:
  Procrustes statistics. {\em Journal of the Royal Statistical Society. Series
  B (Methodological)\/} pp. 234--238

\bibitem[\protect\citeauthoryear{Stephens}{Stephens}{2000}]{stephens00a}
Stephens, M. (2000), Bayesian analysis of mixture models with an unknown number
  of components-an alternative to reversible jump methods. {\em Annals of
  Statistics\/} pp. 40--74

\bibitem[\protect\citeauthoryear{Wasserman and Galaskiewicz}{Wasserman and
  Galaskiewicz}{1994}]{wasserman:galaskiewicz1994}
Wasserman, S. and J.~Galaskiewicz (1994), {\em Advances in social network
  analysis: Research in the social and behavioral sciences\/}. Sage
  Publications, Incorporated

\bibitem[\protect\citeauthoryear{Wasserman and Pattison}{Wasserman and
  Pattison}{1996}]{wasserman:pattison1996}
Wasserman, S. and P.~Pattison (1996), Logit models and logistic regressions for
  social networks: I. An introduction to Markov graphs and p*. {\em
  Psychometrika\/} {\bf 61}(3), 401--425

\bibitem[\protect\citeauthoryear{Wyse and Friel}{Wyse and
  Friel}{2012}]{wyse:friel12}
Wyse, J. and N.~Friel (2012), Block clustering with collapsed latent block
  models. {\em Statistics and Computing\/} {\bf 22}(2), 415--428

\bibitem[\protect\citeauthoryear{Zachary}{Zachary}{1977}]{zachary77}
Zachary, W. (1977), An information flow model for conflict and fission in small
  groups. {\em Journal of anthropological research\/} pp. 452--473

\end{thebibliography}

\end{document}